\documentclass[reprint,superscriptaddress,showframe,nofootinbib,aps,prd,longbibliography,noeprint]{revtex4-2}

\newcommand{\mrm}[1]{\mathrm{#1}}
\newcommand{\eqn}[1]{equation~(#1)}
\newcommand{\fg}[1]{Fig.~#1}

\usepackage{graphicx}% Include figure files
\usepackage{dcolumn}% Align table columns on decimal point
\usepackage{bm}% bold math
\usepackage{orcidlink}
\usepackage{hyperref}
\usepackage{amsmath}
\usepackage{amssymb}
\begin{document}

\preprint{}

\title{Astrophysical systematics in kinematic lensing: quantifying an intrinsic alignment analog}

\author{Yu-Hsiu Huang\orcidlink{0000-0002-4982-0208}}
\email{yhhuang@arizona.edu}
\affiliation{Department of Astronomy/Steward Observatory, University of Arizona, Tucson, Arizona 85721, USA}

\author{Elisabeth Krause\orcidlink{0000-0001-8356-2014}}
\affiliation{Department of Astronomy/Steward Observatory, University of Arizona, Tucson, Arizona 85721, USA}
\affiliation{Department of Physics, University of Arizona, Tucson, Arizona 85721, USA}

\author{Jiachuan Xu\orcidlink{0000-0003-0871-8941}}
\affiliation{Department of Astronomy/Steward Observatory, University of Arizona, Tucson, Arizona 85721, USA}

\author{Tim Eifler\orcidlink{0000-0002-1894-3301}}
\affiliation{Department of Astronomy/Steward Observatory, University of Arizona, Tucson, Arizona 85721, USA}
\affiliation{Department of Physics, University of Arizona, Tucson, Arizona 85721, USA}

\author{Pranjal R. S.\orcidlink{0000-0003-3714-2574}}
\affiliation{Department of Astronomy/Steward Observatory, University of Arizona, Tucson, Arizona 85721, USA}

\author{Eric Huff\orcidlink{0000-0002-9378-3424}}
\affiliation{Jet Propulsion Laboratory, California Institute of Technology, Pasadena, CA 91109, USA}

\date{\today}

\begin{abstract}
Kinematic lensing (KL) is a new weak lensing technique that reduces shape noise for disk galaxies by including spectroscopically measured galaxy kinematics in addition to photometrically measured galaxy shapes. Since KL utilizes the Tully-Fisher relation, any correlation of this relation with the local environment may bias the cosmological interpretation. For the first time, we explore such a Tully-Fisher environmental dependence (TED) effect as a potential astrophysical systematic for KL. Our derivation of the TED systematic can be described in a similar analytical form as intrinsic alignment for traditional weak lensing. We demonstrate analytically that TED only impacts KL if intrinsic alignment for disk galaxies is nonzero. We further use IllustrisTNG simulations to quantify the TED effect. Our two-point correlation measurements do not yield any additional coherent signals that would indicate a systematic bias on KL, within the uncertainties set by the simulation volume. 
\end{abstract}

% \keywords{Suggested keywords}

\maketitle

%\tableofcontents

\section{Introduction} \label{sec:intro}

Weak gravitational lensing (WL) is the deflection in photon paths due to the inhomogeneous large-scale cosmic matter distribution, giving rise to percent-level distortions in galaxy shapes. Since WL probes the integrated line-of-sight matter distribution without any assumption on mass-to-light ratios, it provides a direct measure of the geometric structure and the growth rate of the universe \citep[see][for a review]{Kilbinger2015:review}. 

Over the past decade, WL has emerged as one of the most promising probes for Stage-III photometric surveys, such as the Dark Energy Survey (DES \footnote{\url{https://www.darkenergysurvey.org}}), 
the Kilo-Degree Survey (KiDS\footnote{\url{http://www.astro-wise.org/projects/KIDS/}}), 
and the Hyper Suprime Cam Subaru Strategic Program (HSC\footnote{\url{http://www.naoj.org/Projects/HSC/HSCProject.html}}).
These surveys put percent-level constraints on the parameter \(S_8\equiv \sigma_8 (\Omega_\mrm{m}/0.3)^{0.5}\) \citep{Amon2022:DESY3, Secco2022:DESY3, Asgari2021:KiDS, Dalal2023:HSCY3, Li2023:HSCY3}, with \(\sigma_8\) the amplitude of the density fluctuations and \(\Omega_\mrm{m}\) the present matter density, and \(35\%\)-level on the dark energy equation of state \(w_0\) solely using WL.  Therefore, WL is one of the core cosmological probes for the next-generation ground-based survey, 
the \textit{Vera C. Rubin Observatory} (LSST\footnote{\url{https://www.lsst.org}}), 
and the space-based missions, \textit{Nancy Grace Roman Space Telescope}\footnote{\url{https://roman.gsfc.nasa.gov}} and 
\textit{Euclid}\footnote{\url{https://sci.esa.int/web/euclid}}. These future surveys will significantly reduce the statistical uncertainties, allowing for powerful constraints on the nature of dark energy \citep[e.g.][]{Amendola2018:Euclid, LSST2018:DESC, Eifler2021:Roman}. 

The dominant statistical uncertainty in WL stems from the unknown intrinsic galaxy shapes. In the WL regime, the observed galaxy ellipticity is \(\hat\epsilon^\mrm{obs} \approx \epsilon^\mrm{int} + \gamma\), which is a combination of both the intrinsic galaxy ellipticity \(\epsilon^\mrm{int}\) and the shear \(\gamma\). Observationally, \(\epsilon^\mrm{int}\) and \(\gamma\) are degenerate and the shear measurement precision is limited by the intrinsic ellipticity dispersion \(\sigma_\epsilon\). 
The intrinsic galaxy shapes correlate with the large-scale tidal field, leading to an astrophysical systematic effect called intrinsic alignment (IA). At the two-point statistic level, IA introduces an additional coherent signal between galaxy intrinsic shapes: (1) the so-called II term that describes correlations of intrinsic shapes of galaxies at the same redshift ($\langle \epsilon^\mrm{int}\epsilon^\mrm{int} \rangle$) and (2) the so-called GI term that correlates the intrinsic shape of foreground galaxies with the lensing signal of a background galaxy ($\langle \epsilon^\mrm{int}\gamma \rangle$). Both signals could bias the interpretation of WL measurements significantly if they are not appropriately modeled \citep{Troxel2015:IA, Krause2016:IA}. 

One method to reduce shape noise is to obtain additional information from resolved galaxy kinematics measurements. A shear at $45^\circ$ from a galaxy's major axis, so-called $\gamma_\times$, causes a misalignment between photometric and kinematic minor axes. The measured rotation velocity along the photometric minor axis can thus be used to constrain one shear component \citep[e.g.][]{Blain2002:2DspecWL, Wittman2021:KL}.
\citet{Gurri2020:firstKL} have adopted this idea as a pioneering measurement of galaxy-galaxy lensing with 18 low-redshift galaxies. 

\citet{Huff2013:KL} proposed kinematic lensing (KL) as a technique to obtain the second shear component from galaxy kinematics by including the Tully-Fisher relation \citep[][hereafter TF]{TF1977} as a prior. With this empirical scaling relation, they predict a galaxy's 3D rotational velocity from its luminosity and compare this value with the spectroscopic measurement of the line-of-sight component of the rotational velocity. In the absence of other systematics, one can infer the disk inclination from the difference between the TF prediction and the measurement. This idea has been explored recently by \citet{RS2023:realistic} and \citet{Xu2023:Roman}. Their results suggest that the KL shape noise, \(\sigma^\mrm{KL}_\varepsilon\), falls in the range of 0.022 -- 0.041, an order of magnitude smaller than traditional WL shape noise.

Since KL infers the unlensed (potentially aligned) galaxy shape, the KL shear measurement is immune to IA. However, KL may still be affected by astrophysical systematics similar to IA, if a galaxy's deviation from the mean TF relation is correlated with the environment. We call this hypothetical effect the Tully-Fisher environmental dependence (TED) systematic. A related systematic for weak lensing magnification measurements using the fundamental plane, in the form of a correlation of the size with the environment, has been found in the Horizon-AGN simulation for both spirals and ellipticals \citep{Johnston2023:size}. 

In the context of galaxy formation, the environmental dependence on the TF relation has been studied extensively in observations \citep[e.g.][]{Pelliccia:2019, Perez-Mart:2021, Abril-Melgarejo:2021, Mercier:2022}. There are no conclusive results however, mostly due to the uncertainties of the sample selection, kinematic modeling, and assumptions of galaxy properties. 

This work aims to explore the TED systematic in kinematic lensing analytically and with hydrodynamical simulations. We start with an overview over the KL measurement basics and derive an expression for the TED systematic in Section \ref{sec:theory}. We describe our disk galaxy sample selection and the measurements from simulations in Section \ref{sec:numeric}. We present and discuss our results in Section \ref{sec:result}, \ref{sec:gal-evo}, and \ref{sec:discuss} and conclude in Section \ref{sec:conclusion}. 

\section{Theoretical Background} \label{sec:theory}
Throughout this paper, we denote scalars in regular font and vectors in bold, \(\epsilon\) for complex-value ellipticities and \(\varepsilon\) for the amplitude and scalar component. We refer to estimators with the hat and true values without the hat. 

\subsection{From the Tully-Fisher to the shear estimator}
For a circular disk with an edge-on aspect ratio \(q_z\), the intrinsic galaxy ellipticity at a given disk inclination \(i\) is defined as
\begin{equation} \label{eq:eint}
    \varepsilon^\mrm{int} = \frac{1-\sqrt{1-(1-q_z^2)\sin^2i}}{1+\sqrt{1-(1-q_z^2)\sin^2i}}.
\end{equation} 
If we further assume the rotation axis of the disk is to be perpendicular to the disk, one can measure \(i\) from the ratio of the line-of-sight velocity at the photometric major axis \(v_\mrm{major}\) and the 3D circular rotational velocity \(v_\mrm{circ}\) through
\begin{equation} \label{eq:sini-int}
    \sin{i} = \frac{v_\mrm{major}}{v_\mrm{circ}}.
\end{equation}
While \(v_\mrm{circ}\) is not observable, it can be estimated from the TF relation and the broad-band photometry \(M_\mrm{B}\)
\begin{equation} \label{eq:TFR}
    \log \hat{v}_\mrm{circ} = \log v_\mrm{TF} = b(M_\mrm{B}-M_\mrm{p}) + a,
\end{equation}
where \(a\) is the zero-point, \(b\) is the slope, and \(M_{\rm p}\) is the pivoting magnitude. The TF relation is typically assumed to have log-normal intrinsic scatter \(\sigma_\mrm{TF}\).

However, if the galaxy intrinsically deviates from the TF relation, the aforementioned shape estimation will be biased. We denote this intrinsic deviation from the TF relation for an individual galaxy as
\begin{equation} \label{eq:TFdelta}
    \Delta_\mrm{TF} \equiv \frac{v_\mrm{circ} - v_\mrm{TF}}{v_\mrm{TF}}.
\end{equation}

\begin{figure}[hb]
    \centering
    \includegraphics[width=\columnwidth]{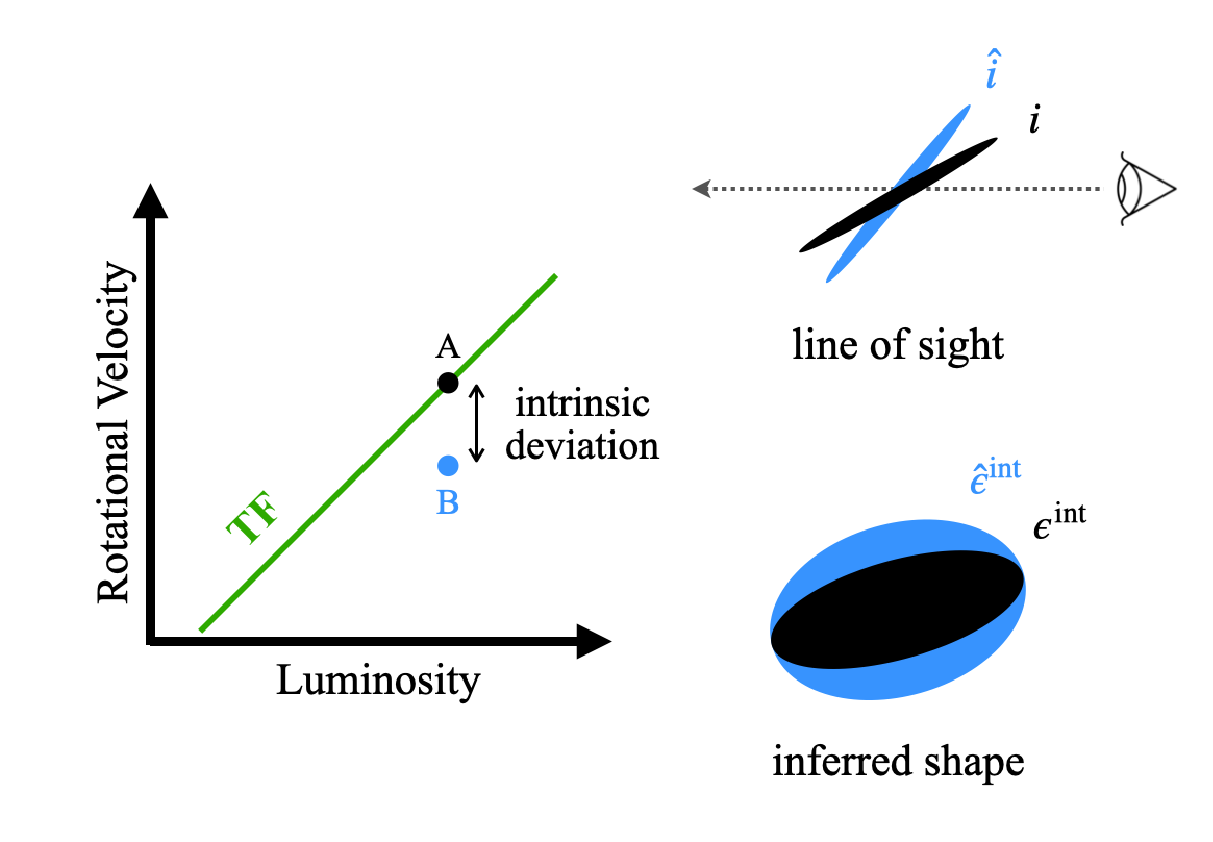}
    \caption{Schematic illustration of how a galaxy's deviation from the TF relation (parametrized by \(\Delta_\mrm{TF}\)) impacts the intrinsic shape estimator\footnote{This figure uses an image created by \href{https://www.flaticon.com/authors/uniconlabs}{Uniconlabs}}. \textit{Left}: The green line shows the TF relation in 3D. Case A (black) corresponds to a galaxy following the TF relation, while case B (blue) illustrates a non-zero \(\Delta_\mrm{TF}\). \textit{Right}: Given a line of sight shown as the dotted line and the galaxy's inclination \(i\), the black part of the figure represents the inclination and the shape of case A; the blue part represents the biased estimates of the inclination and the shape in case B. }
    \label{fig:eTED_concept}
\end{figure}

A conceptual illustration is shown in \fg{\ref{fig:eTED_concept}}. The galaxies in Case A and Case B have the same inclination. Case A does not have the intrinsic offset, i.e. \(\Delta_\mrm{TF}=0\). We can know \(i\) by replacing \(v_\mrm{circ}\) with \(v_\mrm{TF}\) in \eqn{\ref{eq:sini-int}} and therefore infer \(\epsilon^\mrm{int}\). For case B, the galaxy's rotation velocity \(v_\mrm{circ}\) is offseted from the TF prediction \(v_\mrm{TF}\) by \(\Delta_\mrm{TF}\), which results in a smaller line-of-sight velocity. To compensate for that, the inferred inclination \(\hat i\) is biased high, thus the inferred shape \(\hat\epsilon^\mrm{int}\) (the blue ellipse) is different from \(\epsilon^\mrm{int}\) (the black ellipse). 

We first analyze the relation between \(\Delta_\mrm{TF}\) and the difference between the black and the blue inferred shape. Equation (\ref{eq:sini-int}), \eqn{\ref{eq:TFR}}, and \eqn{\ref{eq:TFdelta}} together give the inferred inclination \(\sin \hat i = \frac{v_\mrm{major}}{v_\mrm{TF}} = \sin i \,(1+\Delta_\mrm{TF})\). Due to the small intrinsic scatter of the TF relation, we assume \(\Delta_\mrm{TF}\) to be small and linearize its effect on the estimated ellipticity
\begin{align}
    \hat{\varepsilon}^\mrm{int} 
    &= \frac{1-\sqrt{1-(1-q_z^2)\sin^2\hat{i}}}{1+\sqrt{1-(1-q_z^2)\sin^2\hat{i}}} \nonumber \\
    &= \frac{1-\sqrt{1-(1-q_z^2)\sin{^2i}\, (1+\Delta_\mrm{TF})^{2}}}{1+\sqrt{1-(1-q_z^2)\sin{^2i}\, (1+\Delta_\mrm{TF})^{2}}} \nonumber \\
    &= \varepsilon^\mrm{int} + \varepsilon^\mrm{int} \frac{2\,\Delta_\mrm{TF}}{\sqrt{1-(1-q_z^2)\sin{^2i}}} + \mathcal{O}(\Delta_\mrm{TF}^2). \label{eq:eint-perturb}
\end{align}
Hence, the scalar ellipticity induced by the intrinsic scatter in the TF relation is 
\begin{equation} \label{eq:eTF}
    \varepsilon^\mrm{TED} \approx \varepsilon^\mrm{int} \frac{2\,\Delta_\mrm{TF}}{\sqrt{1-(1-q_z^2)\sin{^2i}}}.
\end{equation}
In addition, the right panel of \fg{\ref{fig:eTED_concept}} already shows that \(\Delta_\mrm{TF}\) changes only the ellipticity but not the position angle, meaning that \(\Delta_\mrm{TF}\) only impacts the ellipticity component along the galaxy's major axis. Hence, the corresponding complex ellipticity reads
\begin{equation}
    \epsilon^\mrm{TED} = \varepsilon^\mrm{TED} e^{i2\varphi}
\end{equation}
with \(\varphi\) being the galaxy's position angle in the source plane.

We further rewrite the estimated intrinsic ellipticity to make the IA contribution explicit,
\begin{equation}
    \hat{\epsilon}^\mrm{int} \approx \epsilon^\mrm{int} + \epsilon^\mrm{TED} \approx \epsilon^\mrm{o} + \epsilon^\mrm{IA} + \epsilon^\mrm{TED},
\end{equation}
where \(\epsilon^\mrm{o}\) is the stochastic intrinsic ellipticity (without IA) and \(\epsilon^\mrm{IA}\) is the IA contribution. On the other hand, the observed ellipticity measured from the galaxy image is \(\hat{\epsilon}^\mrm{obs} \approx \epsilon^\mrm{int} + \gamma \). Since KL measures the unlensed galaxy orientation that includes any alignment component, the KL shear estimator for an individual galaxy is independent of \(\epsilon^\mrm{int}\)
\begin{equation} \label{eq:shear-est}
    \hat{\gamma} \equiv \hat{\epsilon}^\mrm{obs}-\hat{\epsilon}^\mrm{int} \approx \gamma + \epsilon^\mrm{TED}.
\end{equation}
We see that \(\epsilon^\mrm{TED}\) appears as an extra astrophysical component in addition to the true shear \(\gamma\) in the KL shear estimator. 

\subsection{A potential astrophysical systematic for KL}
We calculate the contribution from \(\epsilon^\mrm{TED}\) to the shear two-point correlation functions \(\xi_\pm\) by inserting \eqn{\ref{eq:shear-est}} into the two-point estimator
\begin{align} 
    \xi_\pm (r_\perp)   
    &= \langle \hat{\gamma}_{j,t} \hat{\gamma}_{k,t} \rangle \pm \langle \cdots \rangle_{(\times)} \nonumber\\
    &= \langle (\gamma + \varepsilon^\mrm{TED})_{j,t} (\gamma + \varepsilon^\mrm{TED})_{k,t} \rangle \pm \langle \cdots \rangle_{(\times)} \nonumber\\
    &= \langle \gamma_{j,t} \gamma_{k,t}\rangle \pm \langle \gamma_{\times,j} \gamma_{\times,k} \rangle  + \nonumber \\
    &\phantom{={}} \underset{\mrm{II\,analog}}{\underline{\langle\varepsilon^{\mrm{TED}}_{j,t} \varepsilon^{\mrm{TED}}_{k,t}\rangle }} + 
    \underset{\mrm{GI\,analog}}{\langle \underline{\gamma_{j,t} \varepsilon^{\mrm{TED}}_{k,t} + \varepsilon^{\mrm{TED}}_{j,t} \gamma_{k,t}\rangle}}
    \pm \langle\cdots\rangle_{(\times)} \nonumber \\
    &= \xi^\gamma_\pm (r_\perp)  + \xi^\mrm{TED}_\pm (r_\perp). \label{eq:CF}
\end{align}
Here \(\mathbf{r}_\perp\) is the galaxy coordinate in the plane of a projected map, the ensemble average is over galaxy pairs \(j, k\) satisfying \(\vert \mathbf{r}_{\perp,j}-\mathbf{r}_{\perp, k}\vert = r_\perp\), and \(\langle\cdots\rangle_{(\times)}\) repeats the previous ensemble average expression substituting \(t\) by \(\times\). 
We can read off that if \(\xi^\mrm{TED}_\pm\) is not zero, the TED contamination is analog to II (\(\langle \epsilon^\mrm{TED} \epsilon^\mrm{TED}\rangle\)) and GI (\(\langle \gamma \epsilon^\mrm{TED} \rangle \)) terms.

We can now analyze the conditions under which GI or II analog exist. For the GI analog, we can rewrite it by substituting \(\varepsilon^\mrm{TED}\) with \eqn{\ref{eq:eTF}} 
\begin{equation} \label{eq:GI}
    \langle \gamma_{j,t} \varepsilon^\mrm{TED}_{k,t}\; \rangle = \left\langle \gamma_{j,t} \;\varepsilon^\mrm{int}_{k,t} \frac{2\Delta_{\mrm{TF},k}}{\sqrt{1-(1-q_z^2)\sin^2 i_k}} \right\rangle.
\end{equation}
This expression shows that the galaxy ensemble average is nonzero only if both \(\Delta_\mrm{TF}\) and \(\varepsilon^\mrm{int}_t\) are spatially correlated with the shear, i.e., the integrated density field. The former may be sourced by an environmental dependence of the TF relation; the latter requires disk galaxy to be intrinsically aligned. 

Similarly, we can insert \eqn{\ref{eq:eTF}} into the expression for the II analog in \eqn{\ref{eq:CF}}. After simplification, we see that the result can only be nonzero if \(\langle \epsilon^\mrm{int}_j \Delta_{\mrm{TF},k}\rangle \) or the product \(\langle \epsilon^\mrm{int}_j \epsilon^\mrm{int}_k \rangle \langle \Delta_{\mrm{TF},j} \Delta_{\mrm{TF},k} \rangle \) do not vanish. For the latter, \(\langle \epsilon^\mrm{int}_j \epsilon^\mrm{int}_k \rangle\) is the II term of IA. 
The former expression would be sourced by the GI term of IA (\(\langle \epsilon^\mrm{int}_j \delta_{\mrm{m},k}\rangle \)), unless \(\Delta_\mrm{TF}\) was uncorrelated with \(\delta_{\mrm m}\). However, it is hard to imagine a non-zero correlation between $\Delta_\mrm{TF}$ and $\epsilon^\mrm{int}$ without \(\Delta_\mrm{TF}\) depending on the local environment.

Hence, both the GI and the II analogs require the existence of IA. On large scales, IA of disk galaxies is expected to be dominated by tidal torquing, which is perturbatively suppressed and has not been detected \citep[e.g.][]{Blazek2019:IA, Tonegawa2018:IA, Samuroff2023:IA}. However, on small scales, IA may arise from other environmental processes than tidal torquing and thus induce the TED systematic. \citet{To2024:smallscale} illustrate the sensitivity of cosmic shear to small-scale matter correlation functions, which implies the importance of understanding small-scale systematics. To accurately interpret KL cosmic shear measurement, in this paper, we test for the signature of TED systematic on \(\sim 10\, h^{-1}\mrm{Mpc}\) scales. 

\section{Measuring TED in the TNG simulation} \label{sec:numeric}
While one can read off from \eqn{\ref{eq:GI}} that TED systematic contributions to cosmic shear will be suppressed in the perturbative regime, non-linear modeling is required to study the TED systematic on smaller physical scales. Hence, we measure the TED systematic using hydrodynamic simulations. Below, we briefly introduce the simulation used in this work and describe our procedure for quantifying the TED systematic in simulations. All the position vectors \(\mathbf{d}_j\) are relative to the galaxy's center, defined by the minimal potential, and all the velocity vectors \(\mathbf{v}_j\) are relative to the galaxy's bulk motion.

\subsection{IllustrisTNG} \label{sec:TNG}
\textit{The Next Generation Illustris Simulations}\footnote{\url{https://www.tng-project.org}} \citep[IllustrisTNG, hereafter TNG;][]{Nelson2019:TNGsum, Piilepich2018:TNGstar, Nelson2018:TNGcolor, Naiman2018:TNGchem, Springel2018:TNGcluster, Marinacci2018:TNGmag} are a suite of 18 state-of-art hydrodynamic simulations with different volumes and resolutions. The subgrid physics implemented in TNG broadly reproduces the observed galaxy properties, including color distribution and scaling relations \citep[e.g.][]{Nelson2018:TNGcolor, Piilepich2018:TNGstar}. In this work, we employ the TNG100-1 simulation box, which evolves a side-length 75 \(h^{-1}\)Mpc periodic box with a resolution of \(5.1\times 10^6\,h^{-1}\mrm{M_\odot}\) for dark matter and \(9.4\times 10^5\,h^{-1}\mrm{M_\odot}\) for baryonic particles from $z=127$ to the present day, to maximize the sample size while capturing the kinematic features of galaxies. Throughout this work, we adopt the \emph{Planck 2015} cosmology \citep{Planck2016} of TNG. 

\subsection{Sample selection} \label{sec:sample}

KL targets rotation-supported galaxies with particle motions dominated by the ordered rotation. Hence, we identify a disk galaxy by the ratio of the rotational energy over the total kinematic energy in stellar particles, denoted as \(\kappa_\mrm{rot}\)
\begin{equation}
    \kappa_\mrm{rot}\equiv \frac{\sum\limits_{j} m^\star_j \left\vert\mathbf{d}_j\times \mathbf{v}_j\right\vert^2}{\sum\limits_j m^\star_j v_j^2},
\end{equation}
where \(m^\star_j\) is the stellar particle mass, and the sum runs over all particles within twice the half-mass radius to avoid exterior structure. We consider galaxies with \(\kappa_{\rm rot} > 0.5\) to be rotation-dominated \citep{Sales2012:Kappa, Rodriguez2017:Kappa}.

The second criterion is star formation rate, as disk galaxies are typically star-forming. A common way to separate star-forming and quiescent galaxies in simulations is to set a threshold on the specific star-formation rate \(\rm sSFR \equiv SFR/M_\star\), which reflects the intrinsic color of galaxies. We adopt \(\rm sSFR \geq 0.04 \,Gyr^{-1}\) to separate the star-forming and the quiescent galaxies in TNG \citep{Zjupa2022:selection}. 

We also account for numerical limitations in the galaxy selection. The mass resolution substantially affects kinematic features, such as disk height and ratio between ordered and disordered motions, of any simulated galaxies. Since KL measures the galaxy rotation curve, we need a well-resolved sample to provide accurate kinematic features. As suggested by \citet{Pillepich2019:TNGdisk}, we limit our sample to have at least 1000 stellar particles \(N_\star\) and total mass \(M\) larger than \(10^{9}\,\rm M_\odot\).

In short, our selection criteria are
\begin{enumerate}
    \item \(\kappa_{\rm rot} > 0.5\),
    \item \(\rm sSFR \geq 0.04\,Gyr^{-1}\),
    \item \(N_\star \geq 1000\) and \(M \geq 10^{9}\,\rm M_\odot\).
\end{enumerate}

\subsection{Velocity measurements}
Observationally, disk galaxy kinematics are measured from spectra of emission lines. Since the typical choice of emission lines, for example, \(\mrm{H}\alpha\), [OII], and [OIII], are related to star formation, we measure kinematics from star-forming gas particles, weighting the velocities by each particle's star-formation rate. This weighting mimics the emission-line strength by adopting the canonical relation between the emission line and star formation rate \citep{Kennicutt:1998}.

To construct a galaxy's rotation curve in TNG, we first define the rotational axis by the normalized angular momentum of the gas particles
\begin{equation} \label{eq:angular}
    \hat{\mathbf{L}} = \frac{\sum\limits_{j} m^\mrm{g}_j \mathbf{d}_j\times \mathbf{v}_j}{\left\vert\sum\limits_{j} m^\mrm{g}_j \mathbf{d}_j\times \mathbf{v}_j\right\vert},
\end{equation}
with $m^\mrm{g}_j$ denoting the gas-particle mass, and the sum includes particles within twice the stellar half-mass radius (\(R_{\star,1/2}\)) to exclude extended non-disk structures in the outer regions. For the same reason, we limit the distance to the disk by \(0.5R_{\star,1/2}\) when we calculate the rotation curve. We bin the particles by their distance relative to \(R_{\star,1/2}\) into 20 radial bins. We take a weighted average in each bin to obtain the rotation curve. Finally, we measure the circular velocity \(v_\mrm{circ}\) at the maximum point of the rotation curve. 

\begin{figure}
    \centering
    \includegraphics[width=\columnwidth]{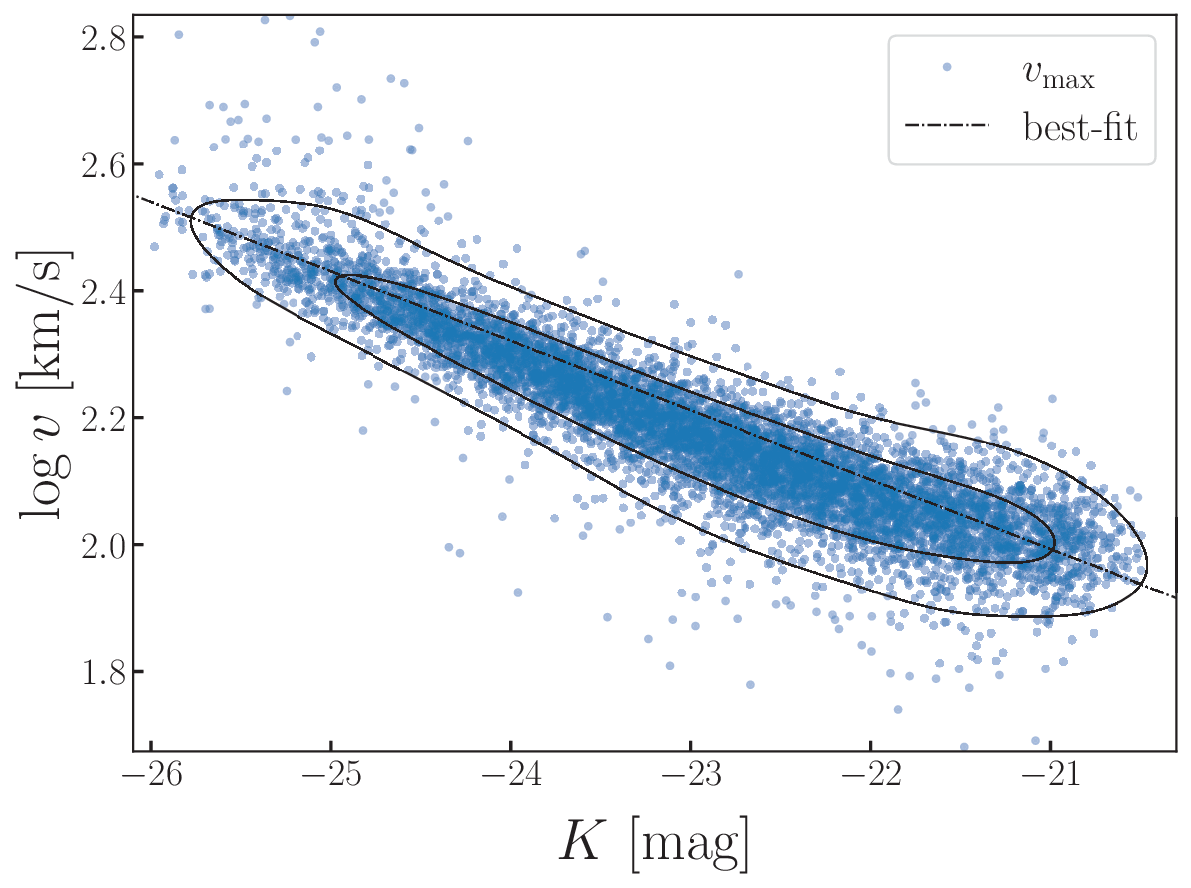}
    \caption{The TF relation of the TNG disk galaxy sample is defined in Section \ref{sec:sample}. The blue dots are our measurements. The dashed-dotted line gives the best-fit TF relation for the \(v_\mrm{circ}\). The contours show the density distribution of the blue dots of 68\% and 95\%.\label{fig:vmax_v25}}
\end{figure}

We obtain the TF relation by fitting \eqn{\ref{eq:TFR}} to the rotation velocity and the \textit{K}-band photometry with three parameters: the zero-point \(a\), the slope \(b\), and the intrinsic scatter \(\sigma_\mrm{TF}\). 
We assume a log-normal distribution for the velocities at fixed magnitude. The variance of the distribution is given by the intrinsic scatter and the measurement uncertainty of the rotation velocity. We quantify these via the standard deviation of the mean velocity. The effective variance for each galaxy is calculated as \(\sigma_{*,j}^2 = \sigma_\mrm{TF}^2 + \sigma_j^2\), where  \(\sigma_j\) is the measurement uncertainty for each galaxy. These lead to the likelihood function
\begin{equation} \label{eq:likelihood}
    \ln\mathcal{L} = -\frac{1}{2}\sum_{j} \frac{\left(\log v_{\mrm{circ}, j} - \log v_{\mrm{TF},j}\right)^2}{\sigma_{*,j}^2} + C
\end{equation}
where \(C\) is a constant. Fitting to the TNG galaxies yields the TF parameters of \(a=2.1882\pm 0.0008\), \(b=-0.1065\pm0.0006\), and \(\sigma_\mrm{TF}=0.0388\pm 0.0006\) dex. The best-fit relation is shown in \fg{\ref{fig:vmax_v25}} with all the measured data from TNG. We then estimate \(\Delta_\mrm{TF}\) using \eqn{\ref{eq:TFdelta}} and \(\epsilon^\mrm{TED}\) via \eqn{\ref{eq:eTF}}. 

\subsection{Ellipticity measurements}
We decompose the projected intrinsic ellipticity \(\epsilon^\mrm{int}\) into the amplitude \(\varepsilon^\mrm{int}\) and the position angle \(\varphi\). For a perfect circular disk with angular momentum perpendicular to the disk, we determine these two components by the projection of the normalized angular momentum from \eqn{\ref{eq:angular}} \citep[e.g.][]{Crittenden:2001ellip}. 

For a projection along the \(z\) direction, the galaxy's inclination is 
\begin{equation}
    i = \cos^{-1} \hat{L}_z,
\end{equation}
from which we obtain \(\varepsilon^\mrm{int}\) through \eqn{\ref{eq:eint}}. Similarly, \(\varphi\) is determined by the projected components \(\hat{L}_x\) and \(\hat{L}_y\),
\begin{equation}
    \varphi = \tan^{-1}\left(\frac{\hat{L}_y}{\hat{L}_x}\right).
\end{equation}
Together, these determine the projected ellipticity 
\begin{equation}
    \epsilon^\mrm{int} = \varepsilon^\mrm{int} e^{i2\varphi} = \varepsilon^\mrm{int}_1 + i\varepsilon^\mrm{int}_2.
\end{equation}

\subsection{Characterizing the environment}\label{sec:env}
We consider two environmental indicators in the simulations: the matter overdensity and the tidal anisotropy. We first construct the matter density field \(\rho_\mrm{m}(\mathbf{x})\) by assigning particles to a \(256^3\) mesh via Cloud-in-Cell algorithm and smooth it by a Gaussian kernel with smoothing length \(R_\mrm{s}\) to aid with numerical stability and to isolate effects of different physical scales. The overdensity at each grid point is given by
\begin{equation}
    \delta_\mrm{m}(\mathbf{x}) = \frac{\rho_\mrm{m}(\mathbf{x})}{\langle \rho_\mrm{m} \rangle} - 1,
\end{equation}
where \(\langle \rho_\mrm{m}\rangle\) is the mean matter density, and the tidal tensor is defined as
\begin{equation}
    T_{jk}(\mathbf{x}) = \frac{\partial^2\Phi(\mathbf{x})}{\partial x_j\partial x_k}. 
\end{equation}
The gravitational potential \(\Phi(\mathbf{x})\) is computed from \(\rho_\mrm{m}(\mathbf{x})\) through the Poisson equation and \(j,k \in (x,y,z)\) are the spatial coordinate axes. By default, we set \(R_\mrm{s}=1 \,h^{-1}\)Mpc; we will discuss the impact of this choice in Section~\ref{sec:discuss-Rs}. Finally, we use the inverse Could-in-Cell algorithm to interpolate the overdensity and the tidal tensor from the grid to arbitrary positions. 

From \(T_{jk}(\mathbf{x})\), we calculate the eigenvalues \(\lambda_1 \leq \lambda_2 \leq \lambda_3\), and measure the tidal anisotropy \citep{Heavens1988:tensor_shear, Catelan1996:tensor_shear}
\begin{equation} \label{eq:qlambda}
    q^2_\lambda \equiv \frac{1}{2} \left[(\lambda_1-\lambda_2)^2+(\lambda_3-\lambda_2)^2+(\lambda_1-\lambda_3)^2\right].
\end{equation}
\(q_\lambda\) represents the strength of the tidal shear at a given position. Since \(q_\lambda\) is measured from the second-order derivative of the potential field, \(q_\lambda\) may appear redundant to \(\delta_\mrm{m}\) at first glance. However, \citet{Sheth2002:q_diff_from_delta} showed that \(q_\lambda\) and \(\delta_\mrm{m}\) encode complementary physical information and follow different distributions. 

\section{Results: quantifying TED} \label{sec:result}
For TED to become a systematic of the KL shear measurement, both \(\Delta_\mrm{TF}\)  and \(\epsilon^\mrm{TED}\) have to be spatially correlated with the environment (see \eqn{\ref{eq:GI}}). In this section, we use TNG to test for these spatial correlations. 

All the correlation functions in this work are calculated using the python package \texttt{TreeCorr}\footnote{\url{https://github.com/rmjarvis/TreeCorr}} \citep{treecorr}. We estimate the covariances through the Jackknife algorithm implemented in \texttt{TreeCorr}. By default, we measure the correlation function in the projected comoving coordinate and set the default smoothing scales of the overdensity and tidal field anisotropy to \(R_\mrm{s}=1\,h^{-1}\mrm{Mpc}\).

\begin{figure}[hbt]
    \centering
    \includegraphics[width=\columnwidth]{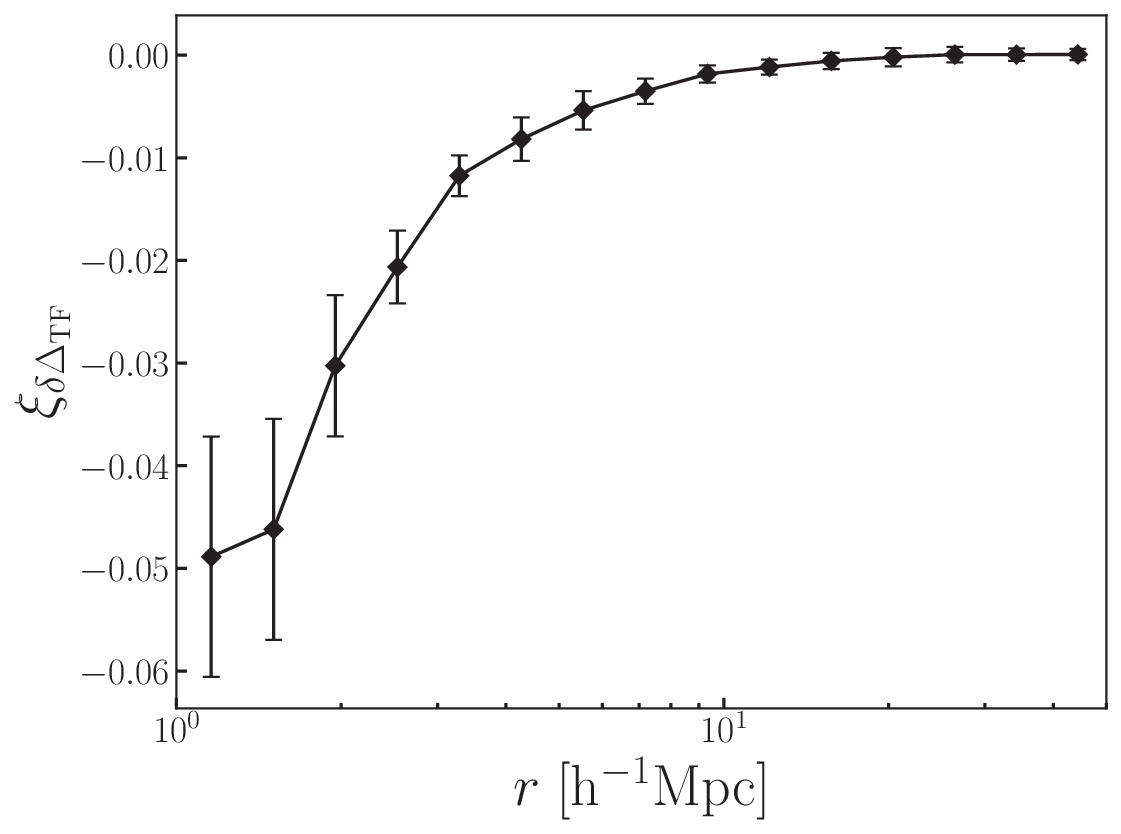}
    \caption{The correlation function between \(\delta_\mrm{m}\) and \(\Delta_\mrm{TF}\). This implies that \(\Delta_\mrm{TF}\) is lower in the denser environment.}
    \label{fig:density-offset}
\end{figure}

We measure TED via the three-dimensional correlation function of \(\delta_\mrm{m}\) and \(\Delta_\mrm{TF}\). \fg{\ref{fig:density-offset}} shows localized negative correlations between the two quantities, and then the function approaches zero as the scale increases. This indicates that TED exists and that \(\Delta_\mrm{TF}\) anti-correlates with \(\delta_\mrm{m}\) at few-Mpc scales.

\begin{figure*}[hbt]
    \centering
    \includegraphics[width=\textwidth]{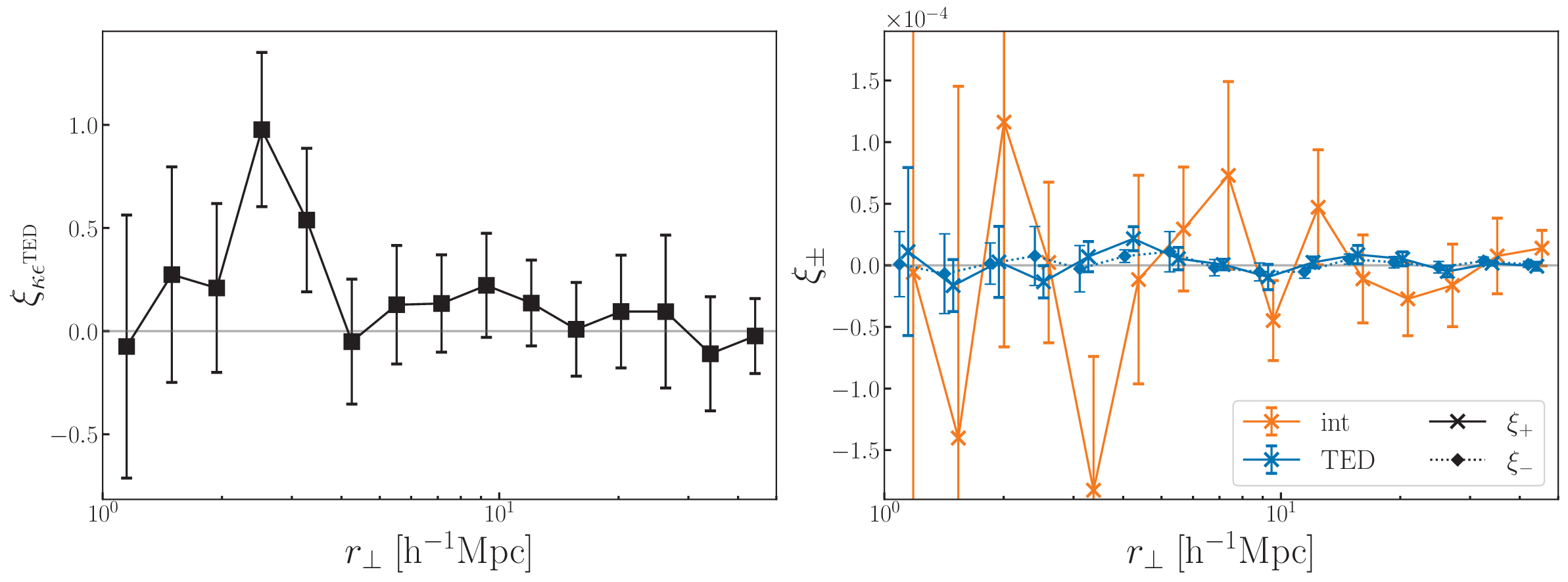}
    \caption{\textit{Left}: the projected cross-correlation function between \(\epsilon^\mrm{TED}\) and the projected matter density \(\kappa\).  \textit{Right}: the projected ellipticity correlation functions of \(\langle\epsilon^\mrm{TED} \epsilon^\mrm{TED}\rangle_\pm\) (blue) and \(\langle \epsilon^\mrm{int} \epsilon^\mrm{int} \rangle_\pm\) (orange). The solid and the dotted lines stand for \(\xi_t\) and \(\xi_-\), respectively. For \(\langle \epsilon^\mrm{int} \epsilon^\mrm{int} \rangle\), we only show the plus term for simplicity.}
    \label{fig:TED-GI-II}
\end{figure*}

While WL analyses measure shears that are the result of line-of-sight projections with long projection lengths, for the purpose of isolating TED systematics, we choose to work on simulation snapshots rather than lightcones. The physical correlation of TED or intrinsic galaxy shape with environment is diluted by projection, and 3D measurements are most discriminating; however, as \(\epsilon^\mrm{TED}\) is defined only in projection, we use plane-parallel projections of the simulation snapshots to measure the GI and II analogs. 

We choose the \textit{z}-axis of the simulation box as the line-of-sight direction and project each galaxy's angular momentum and rotational velocity accordingly to measure \(\epsilon^\mrm{int}\) and \(\varepsilon^\mrm{TED}\) following section \ref{sec:numeric}. We use the \(z=1\) snapshot from TNG and report the results in comoving distance.

For the plane-parallel projection, we define the projected density contrast \(\kappa\) as
\begin{equation}
    \kappa (\mathbf{x}) = \int_0^L \delta_\mrm{m}(\mathbf{x}) dz\,,
\end{equation}
where we integrate \(\delta_\mrm{m}\) along the \(z\)-axis and \(L\) is the box size. Note that this definition differs from the lensing convergence as it does not include any lens efficiency weighting.

\paragraph{GI analogs} 
To quantify TED-induced GI-type contamination, we measure the projected cross-correlation function between \(\epsilon^{\rm{TED}}\) and $\kappa$. The result is consistent with zero given the statistical uncertainty, as shown in the left panel of \fg{\ref{fig:TED-GI-II}}. The reduced $\chi^2$ value for the 20 bins is 0.49, indicating an insignificant correlation between \(\epsilon^\mrm{TED}\) and \(\kappa\). We perform the exact measurement for \(\epsilon^\mrm{TED}\) and the tidal anisotropy \(q_\lambda\), which is also consistent with zero at all separations yielding a \(\chi^2\) value of 0.81 over 20 bins.

\paragraph{II analogs} 
We measure the projected ellipticity correlation functions $\xi_\pm^{\rm{TED}}$ analog to the II terms, shown in the right panel of \fg{\ref{fig:TED-GI-II}}. We adopt a \(\sin i\)-based selection at \(\sin i \leq 0.8\) to the sample since the approximation of \(\varepsilon^\mrm{TED}\) using \eqn{\ref{eq:eTF}} is suitable only for samples with small \(\Delta_\mrm{TF}\) and low \(\sin i\). This selection does not induce a bias as long as the galaxies are oriented randomly. The auto-correlation \(\langle\epsilon^\mrm{TED} \epsilon^\mrm{TED}\rangle_\pm\), illustrated as blue solid and dotted lines, are in agreement with zero even at small scales with \(\chi^2=0.73\) and 0.60, respectively. The figure indicates that we do not find evidence of II analogs for KL. The shape noise of \(\epsilon^\mrm{TED}\), denoted \(\sigma^\mrm{TED}_\varepsilon\), is 0.031, which is comparable to \(\sigma^\mrm{KL}_\varepsilon=0.035\) that \citet{Xu2023:Roman} derive for a potential KL Roman survey. With more sophisticated forward modeling, we can cope with the instability of \eqn{\ref{eq:eTF}} and reduce \(\sigma^\mrm{TF}_\varepsilon\). 
Furthermore, we measure the intrinsic alignment correlations \(\langle \epsilon^\mrm{int} \epsilon^\mrm{int} \rangle_\pm\) shown in orange. Within the statistical uncertainty, the measurement agrees with zero. In short, we do not detect any coherent signal for \(\epsilon^\mrm{TED}\).

\section{Results: TED and galaxy populations} \label{sec:gal-evo}

\begin{figure*}[htb]
    \centering
    \includegraphics[width=\textwidth]{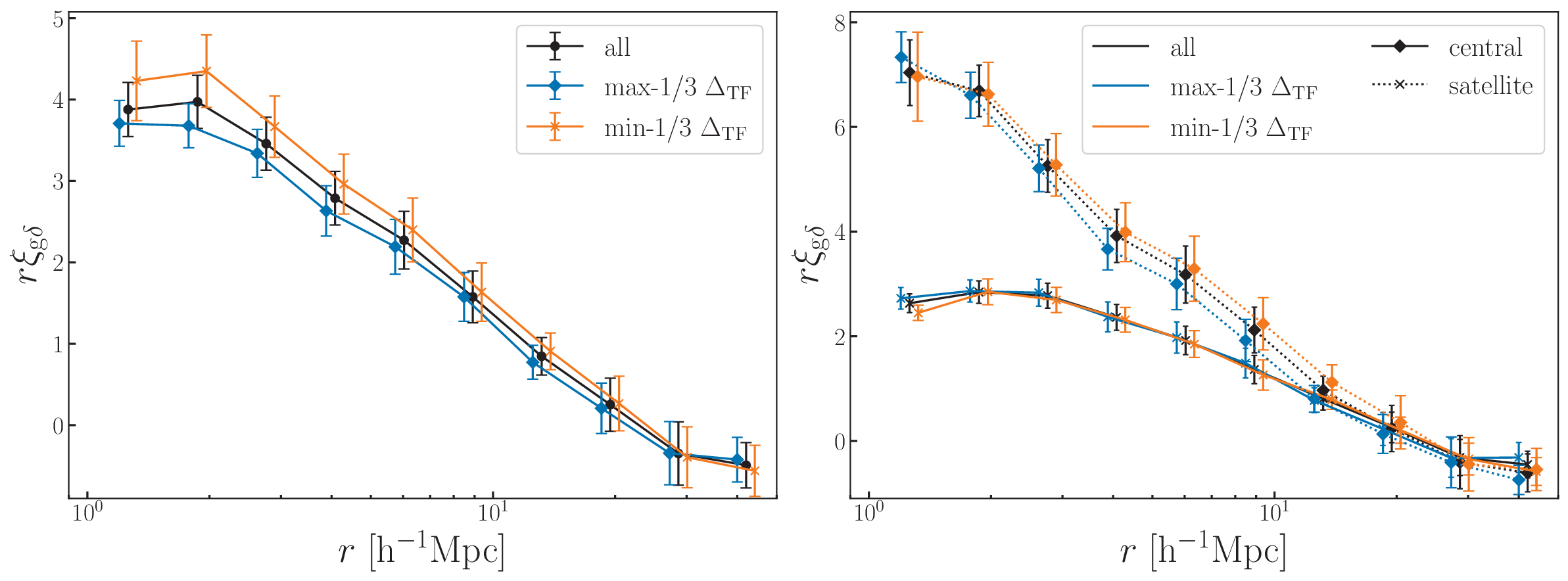}
    \caption{\textit{Left}: the galaxy-matter cross-correlation function for the max-1/3 (blue), the min-1/3 (orange), and the whole ensemble (black). The min-1/3 galaxies reside in denser environments than the ensemble average, while the max-1/3 galaxies live in less dense environments. We have shifted each curve by 0.05 dex in the $x$ axis for better illustration. \textit{Right}: The galaxy-matter cross-correlation function decomposed by the central (solid) and the satellite (dotted) populations. The color code is the same as the left panel. The two groups have consistent central and satellite correlation fuctions.}
    \label{fig:spin0-3D-density-split}
\end{figure*}

Galaxies with different \(\Delta_\mrm{TF}\) may originate from distinct populations, potentially behaving differently in the galaxy-matter correlation and clustering. Consequently, we classify our galaxies that fall into the lower (upper) tercile of the distribution of \(\Delta_\mrm{TF}\) at fixed luminosity as min-1/3 (max-1/3) galaxies. We measure the galaxy-density cross-correlation \(\xi_{\mrm{g}\delta}\) for each group in the left panel of \fg{\ref{fig:spin0-3D-density-split}} showing the three-dimensional galaxy-matter correlation functions. We find an increase in amplitude of \(\xi_{\mrm{g}\delta}\) for the min-1/3 group at \(r\lesssim 10 \,h^{-1}\mrm{Mpc}\) compared to the ensemble average, suggesting that the lower tercile galaxies tend to live in denser environments. The upper tercile galaxies, on the other hand, have suppressed correlation functions in the same regime. The result indicates that there is an anticorrelation between \(\Delta_\mrm{TF}\) and the density of the host environment. 

We further measure each group's satellite fraction \(f_\mrm{sat}\). The value is \(f_\mrm{sat}=0.29\) for the whole sample, 0.22 for the max-1/3 galaxies, and 0.39 for the min-1/3 galaxies. By measuring the galaxy-galaxy correlation functions, we find an apparent deviation of the min-1/3 from the ensemble behavior, indicating the min-1/3 population is more strongly clustered. This behavior is consistent with a higher \(f_\mrm{sat}\) in groups or clusters.

If we denote the galaxy-matter correlation function for satellite and central as \(\xi_{\mrm{g}\delta}^\mrm{sat}\) and \(\xi_{\mrm{g}\delta}^\mrm{cen}\), respectively, then \(\xi_{\mrm{g}\delta}\) for a group of galaxies with satellite fraction \(f_\mrm{sat}\) is \(\xi_{\mrm{g}\delta} = (1-f_\mrm{sat})\xi_{\mrm{g}\delta}^\mrm{cen} + f_\mrm{sat} \xi_{\mrm{g}\delta}^\mrm{sat}\). Therefore, we separately measure \(\xi_{\mrm{g}\delta}^\mrm{sat}\) and \(\xi_{\mrm{g}\delta}^\mrm{cen}\) for each group in the right panel of \fg{\ref{fig:spin0-3D-density-split}}. After the separation, we do not find any clear deviation among different groups in either \(\xi_{\mrm{g}\delta}^\mrm{sat}\) or \(\xi_{\mrm{g}\delta}^\mrm{cen}\). This indicates that \(f_\mrm{sat}\) can explain the behavior \(\xi_{\mrm{g}\delta}\) for different galaxy populations in the left panel. 

We further investigate whether the difference in large-scale clustering amplitude between the two populations (\fg{\ref{fig:spin0-3D-density-split}}) may be a form of assembly bias. We predict mean linear bias of the two populations from the \citet{Tinker2010:bias} halo bias averaged over their respective host halo mass \(\mrm{M_{200c}}\) distributions. This calculation predicts a relative bias between the max-1/3 and the min-1/3 galaxies to be 1.0634, which is statistically consistent with our \(\xi_{\mrm{g}\delta}\) measurements, and thus provides no indication for large-scale assembly bias.

\section{Discussion} \label{sec:discuss}
We have presented the measurements of GI and II analogs on TNG galaxies, which do not show any evidence of TED systematic for KL. We note that these conclusions are subject to the statistical limitation of a 75 \(h^{-1}\mrm{Mpc}\) simulation box and the galaxy formation model implemented in TNG. Increasing the box size will increase the sample size, potentially leading to detection at small separations in \fg{\ref{fig:TED-GI-II}}. The role of the galaxy formation model is much more complicated. Although TNG and other cosmological simulations broadly reproduce the observed galaxy properties, different models still give slightly different predictions on the environmental dependence in terms of strength and correlation among different properties \citep[e.g.][]{Dave2020}. Furthermore, different simulations give different predictions on the intrinsic alignment of spiral galaxies \citep[e.g.][]{Tenneti2016:IA}.

We also emphasize that the measurements presented in \fg{\ref{fig:TED-GI-II}} are only projected along the simulation snapshot, which enhances the significance of any TED correlations. In practice, the projection of angular shear two-point statistics along the lightcone will average out the TED correlation as described in section \ref{sec:result}. Hence, the tests presented here provide a conservative assessment of TED systematics.

Furthermore, our choices during the sample selection and extracting the environment may bias our conclusions. To test the robustness of our results, we vary the definitions of the environments and the choice of sample selection criteria.

\subsection{Definition of environment} \label{sec:discuss-Rs}

Different galaxy formation mechanisms are dominant at different physical scales. Thus, choosing a specific \(R_\mrm{s}\) presumably targets certain processes and smears out other minor effects, leading to biased conclusions. Since this work aims to robustly investigate the possibility of TED leading to a GI analog, it is essential that our conclusion in section \ref{sec:result} is not affected by the choice of \(R_\mrm{s}\). We test the robustness with three different \(R_\mrm{s}\): 0.5, 1.0, and 5.0 \(h^{-1}\)Mpc. We measure the GI analog for each definition and calculate the corresponding \(\chi^2\) value to quantify how significantly the correlation function deviates from the zero.

We do not observe any signal of correlation between \(\epsilon^\mrm{TED}\) and \(\kappa\) at these different \(R_\mrm{s}\). The larger \(R_\mrm{s}\) leads to a smaller correlation function amplitude and smaller uncertainties. Even though the uncertainties shrink as \(R_\mrm{s}\) increases, the reduced \(\chi^2\) suggests that the associate correlation function is still consistent with zero. We repeat the same test on \(q_\lambda\) and find the same conclusion. In both cases, the variation of \(R_\mrm{s}\) does not result in any GI analog.

\subsection{Galaxy selection} \label{sec:discuss-sample}
The sample selection can impact $\Delta_\mrm{TF}$, the alignment, or the clustering of galaxies. We start with the variation in $\Delta_\mrm{TF}$ to look for potential variables of selection criteria worth investigating and then measure the two-point correlation functions to understand the influence on the alignment and the clustering. 

The selection criteria are based on four galaxy properties: $\kappa_\mrm{rot}$, sSFR, $N_\star$, and $\rm M$. We calculate the correlation coefficients between $\Delta_\mrm{TF}$ and the four properties. Most importantly, none of the four properties has a statistically significant correlation with $\Delta_\mrm{TF}$, suggesting that the TF is robust against variation of the aforementioned variables. Among the four, $\Delta_\mrm{TF}$ most strongly correlates with \(\rm M\). $N_\star$ shows the second highest correlation, mainly associated with the tight relation between $N_\star$ and $\mrm{M}$, followed by $\kappa_\mrm{rot}$. sSFR shows the least and almost zero relation to $\Delta_\mrm{TF}$.

In addition to the selection effect on the TED, we also look for its influence on IA. Since massive galaxies generally form earlier, they are more likely aligned by their host dark matter halos than less massive galaxies. On the other hand, a more rotation-dominated system is more affected by the tidal torque. Thus, we investigate the impact on the GI analog in two cases: massive galaxies where \(\mrm{M > 10^{12}M_\odot}\) and the highly rotation-dominated systems where \(\kappa_\mrm{rot} > 0.7\). The reduced \(\chi^2\) values are 0.72 and 1.11, respectively, implying no detection of TED systematic.

\section{Conclusion} \label{sec:conclusion}
KL is a promising technique for probing cosmic structure formation with high statistical precision. It is also insensitive towards observational uncertainties that affect traditional weak lensing, such as shear calibration and photo-z errors. However, if deviations from the TF relation are spatially correlated with large-scale structure, this may induce an IA-like contamination to the KL measurement. This is the first paper to study this astrophysical systematics, termed TED (Tully Fisher environmental dependence), analytically and with state-of-the-art hydrodynamic simulations TNG.

We first show how TED may bias KL analogously to IA in traditional WL by deriving the TED-induced ellipticity \(\epsilon^\mrm{TED}\). Our derivation shows that both TED and intrinsic alignment for disk galaxies on \(\sim 10\,h^{-1}\mrm{Mpc}\) scales must exist for the bias to be non-zero . We further quantify the TED systematic by measuring the GI and II analogs from TNG galaxies. For the GI analogs, we measure the cross-correlation for \(\langle\kappa \epsilon^\mrm{TED}\rangle\) and \(\langle q_\lambda\epsilon^\mrm{TED}\rangle\), respectively. For the II analogs, we measure the auto-correlation \(\langle\epsilon^\mrm{TED} \epsilon^\mrm{TED}\rangle\) and the cross-correlation \(\langle\epsilon^\mrm{TED} \epsilon^\mrm{int}\rangle\). We find the reduced \(\chi^2\) values for both measurements to be consistent with zero within the measurement error, meaning that we do not find any coherent TED systematic that would bias KL measurements. Finally, we demonstrate the robustness of the definition of both \(\kappa\) and \(q_\lambda\) and sample selection. %We conclude that our results are robust to both definitions.

We also report a different type of TED that does not lead to spatial coherent signals in the two-point shear measurement. We find that galaxies rotating more slowly than the TF prediction tend to live in denser environments. We attribute this dependence to the satellite fraction of each population. 

In summary, this work indicates that an environmental dependence of the Tully-Fisher relation does not cause systematic biases for KL. As our results are limited by the statistical power of the TNG100 simulation, future KL analyses should validate these findings with a larger simulation volume to reduce the statistical uncertainties and include realistic mock observations to account for potential systematic biases due to kinematic substructure. The TED systematic can also be tested observationally with KL measurements on nearby well-resolved galaxies, for which we do not expect shear detection but only systematics if they exist.

\section*{Acknowledgements}
The authors thank Andr\'es N. Salcedo for calculating the linear bias and the useful discussion on assembly bias. Y.-H. H. thanks Spencer Everett for helpful discussions and comments.
This work was supported by NASA ROSES grants ADAP 20-ADAP20-0158 and Roman WFS 22-ROMAN22-0016. E. K., Y.-H. H. and P. R. S. were supported in part by the David and Lucile Packard Foundation and an Alfred P. Sloan Research Fellowship. The analyses in this work were carried out using the High Performance Computing (HPC) resources supported by the University of Arizona the 611 Technology and Research Initiative Fund (TRIF), University 612 Information Technology and Service (UITS), and Office for 613 Research, Innovation, and Impact (RDI) and maintained by the UA Research Technologies Department.

\bibliographystyle{apsrev4-2}
\bibliography{main}

\providecommand{\mnras}{Monthly Notices of the Royal Astronomical Society}\providecommand{\araa}{Annual Review of Astronomy and Astrophysics}\providecommand{\apj}{Astrophysical Journal}\providecommand{\apjl}{Astrophysical Journal Letter}\providecommand{\aap}{Astronomy and Astrophysics}\providecommand{\physrep}{Physics Reports}\providecommand{\pasj}{Publications of the ASJ}
\begin{thebibliography}{47}%
\makeatletter
\providecommand \@ifxundefined [1]{%
 \@ifx{#1\undefined}
}%
\providecommand \@ifnum [1]{%
 \ifnum #1\expandafter \@firstoftwo
 \else \expandafter \@secondoftwo
 \fi
}%
\providecommand \@ifx [1]{%
 \ifx #1\expandafter \@firstoftwo
 \else \expandafter \@secondoftwo
 \fi
}%
\providecommand \natexlab [1]{#1}%
\providecommand \enquote  [1]{``#1''}%
\providecommand \bibnamefont  [1]{#1}%
\providecommand \bibfnamefont [1]{#1}%
\providecommand \citenamefont [1]{#1}%
\providecommand \href@noop [0]{\@secondoftwo}%
\providecommand \href [0]{\begingroup \@sanitize@url \@href}%
\providecommand \@href[1]{\@@startlink{#1}\@@href}%
\providecommand \@@href[1]{\endgroup#1\@@endlink}%
\providecommand \@sanitize@url [0]{\catcode `\\12\catcode `\$12\catcode `\&12\catcode `\#12\catcode `\^12\catcode `\_12\catcode `\%12\relax}%
\providecommand \@@startlink[1]{}%
\providecommand \@@endlink[0]{}%
\providecommand \url  [0]{\begingroup\@sanitize@url \@url }%
\providecommand \@url [1]{\endgroup\@href {#1}{\urlprefix }}%
\providecommand \urlprefix  [0]{URL }%
\providecommand \Eprint [0]{\href }%
\providecommand \doibase [0]{https://doi.org/}%
\providecommand \selectlanguage [0]{\@gobble}%
\providecommand \bibinfo  [0]{\@secondoftwo}%
\providecommand \bibfield  [0]{\@secondoftwo}%
\providecommand \translation [1]{[#1]}%
\providecommand \BibitemOpen [0]{}%
\providecommand \bibitemStop [0]{}%
\providecommand \bibitemNoStop [0]{.\EOS\space}%
\providecommand \EOS [0]{\spacefactor3000\relax}%
\providecommand \BibitemShut  [1]{\csname bibitem#1\endcsname}%
\let\auto@bib@innerbib\@empty
%</preamble>
\bibitem [{\citenamefont {{Kilbinger}}(2015)}]{Kilbinger2015:review}%
  \BibitemOpen
  \bibfield  {author} {\bibinfo {author} {\bibfnamefont {M.}~\bibnamefont {{Kilbinger}}},\ }\href {https://doi.org/10.1088/0034-4885/78/8/086901} {\bibfield  {journal} {\bibinfo  {journal} {Reports on Progress in Physics}\ }\textbf {\bibinfo {volume} {78}},\ \bibinfo {eid} {086901} (\bibinfo {year} {2015})},\ \Eprint {https://arxiv.org/abs/1411.0115} {arXiv:1411.0115 [astro-ph.CO]} \BibitemShut {NoStop}%
\bibitem [{\citenamefont {{Amon}}\ \emph {et~al.}(2022)\citenamefont {{Amon}}, \citenamefont {{Gruen}}, \citenamefont {{Troxel}}, \citenamefont {{MacCrann}}, \citenamefont {{Dodelson}},\ and\ \citenamefont {{DES Collaboration}}}]{Amon2022:DESY3}%
  \BibitemOpen
  \bibfield  {author} {\bibinfo {author} {\bibfnamefont {A.}~\bibnamefont {{Amon}}}, \bibinfo {author} {\bibfnamefont {D.}~\bibnamefont {{Gruen}}}, \bibinfo {author} {\bibfnamefont {M.~A.}\ \bibnamefont {{Troxel}}}, \bibinfo {author} {\bibfnamefont {N.}~\bibnamefont {{MacCrann}}}, \bibinfo {author} {\bibfnamefont {S.}~\bibnamefont {{Dodelson}}},\ and\ \bibinfo {author} {\bibnamefont {{DES Collaboration}}},\ }\href {https://doi.org/10.1103/PhysRevD.105.023514} {\bibfield  {journal} {\bibinfo  {journal} {\prd}\ }\textbf {\bibinfo {volume} {105}},\ \bibinfo {eid} {023514} (\bibinfo {year} {2022})},\ \Eprint {https://arxiv.org/abs/2105.13543} {arXiv:2105.13543 [astro-ph.CO]} \BibitemShut {NoStop}%
\bibitem [{\citenamefont {{Secco}}\ \emph {et~al.}(2022)\citenamefont {{Secco}}, \citenamefont {{Samuroff}}, \citenamefont {{Krause}}, \citenamefont {{Jain}}, \citenamefont {{Blazek}},\ and\ \citenamefont {{DES Collaboration}}}]{Secco2022:DESY3}%
  \BibitemOpen
  \bibfield  {author} {\bibinfo {author} {\bibfnamefont {L.~F.}\ \bibnamefont {{Secco}}}, \bibinfo {author} {\bibfnamefont {S.}~\bibnamefont {{Samuroff}}}, \bibinfo {author} {\bibfnamefont {E.}~\bibnamefont {{Krause}}}, \bibinfo {author} {\bibfnamefont {B.}~\bibnamefont {{Jain}}}, \bibinfo {author} {\bibfnamefont {J.}~\bibnamefont {{Blazek}}},\ and\ \bibinfo {author} {\bibnamefont {{DES Collaboration}}},\ }\href {https://doi.org/10.1103/PhysRevD.105.023515} {\bibfield  {journal} {\bibinfo  {journal} {\prd}\ }\textbf {\bibinfo {volume} {105}},\ \bibinfo {eid} {023515} (\bibinfo {year} {2022})},\ \Eprint {https://arxiv.org/abs/2105.13544} {arXiv:2105.13544 [astro-ph.CO]} \BibitemShut {NoStop}%
\bibitem [{\citenamefont {{Asgari}}\ \emph {et~al.}(2021)\citenamefont {{Asgari}}, \citenamefont {{Lin}}, \citenamefont {{Joachimi}}, \citenamefont {{Giblin}}, \citenamefont {{Heymans}},\ and\ \citenamefont {et~al.}}]{Asgari2021:KiDS}%
  \BibitemOpen
  \bibfield  {author} {\bibinfo {author} {\bibfnamefont {M.}~\bibnamefont {{Asgari}}}, \bibinfo {author} {\bibfnamefont {C.-A.}\ \bibnamefont {{Lin}}}, \bibinfo {author} {\bibfnamefont {B.}~\bibnamefont {{Joachimi}}}, \bibinfo {author} {\bibfnamefont {B.}~\bibnamefont {{Giblin}}}, \bibinfo {author} {\bibfnamefont {C.}~\bibnamefont {{Heymans}}},\ and\ \bibinfo {author} {\bibnamefont {et~al.}},\ }\href {https://doi.org/10.1051/0004-6361/202039070} {\bibfield  {journal} {\bibinfo  {journal} {\aap}\ }\textbf {\bibinfo {volume} {645}},\ \bibinfo {eid} {A104} (\bibinfo {year} {2021})},\ \Eprint {https://arxiv.org/abs/2007.15633} {arXiv:2007.15633 [astro-ph.CO]} \BibitemShut {NoStop}%
\bibitem [{\citenamefont {{Dalal}}\ \emph {et~al.}(2023)\citenamefont {{Dalal}}, \citenamefont {{Li}}, \citenamefont {{Nicola}}, \citenamefont {{Zuntz}}, \citenamefont {{Strauss}},\ and\ \citenamefont {et~al.}}]{Dalal2023:HSCY3}%
  \BibitemOpen
  \bibfield  {author} {\bibinfo {author} {\bibfnamefont {R.}~\bibnamefont {{Dalal}}}, \bibinfo {author} {\bibfnamefont {X.}~\bibnamefont {{Li}}}, \bibinfo {author} {\bibfnamefont {A.}~\bibnamefont {{Nicola}}}, \bibinfo {author} {\bibfnamefont {J.}~\bibnamefont {{Zuntz}}}, \bibinfo {author} {\bibfnamefont {M.~A.}\ \bibnamefont {{Strauss}}},\ and\ \bibinfo {author} {\bibnamefont {et~al.}},\ }\href {https://doi.org/10.48550/arXiv.2304.00701} {\bibfield  {journal} {\bibinfo  {journal} {arXiv e-prints}\ ,\ \bibinfo {eid} {arXiv:2304.00701}} (\bibinfo {year} {2023})},\ \Eprint {https://arxiv.org/abs/2304.00701} {arXiv:2304.00701 [astro-ph.CO]} \BibitemShut {NoStop}%
\bibitem [{\citenamefont {{Li}}\ \emph {et~al.}(2023)\citenamefont {{Li}}, \citenamefont {{Zhang}}, \citenamefont {{Sugiyama}}, \citenamefont {{Dalal}}, \citenamefont {{Rau}},\ and\ \citenamefont {et~al.}}]{Li2023:HSCY3}%
  \BibitemOpen
  \bibfield  {author} {\bibinfo {author} {\bibfnamefont {X.}~\bibnamefont {{Li}}}, \bibinfo {author} {\bibfnamefont {T.}~\bibnamefont {{Zhang}}}, \bibinfo {author} {\bibfnamefont {S.}~\bibnamefont {{Sugiyama}}}, \bibinfo {author} {\bibfnamefont {R.}~\bibnamefont {{Dalal}}}, \bibinfo {author} {\bibfnamefont {M.~M.}\ \bibnamefont {{Rau}}},\ and\ \bibinfo {author} {\bibnamefont {et~al.}},\ }\href {https://doi.org/10.48550/arXiv.2304.00702} {\bibfield  {journal} {\bibinfo  {journal} {arXiv e-prints}\ ,\ \bibinfo {eid} {arXiv:2304.00702}} (\bibinfo {year} {2023})},\ \Eprint {https://arxiv.org/abs/2304.00702} {arXiv:2304.00702 [astro-ph.CO]} \BibitemShut {NoStop}%
\bibitem [{\citenamefont {{Amendola}}\ \emph {et~al.}(2018)\citenamefont {{Amendola}}, \citenamefont {{Appleby}}, \citenamefont {{Avgoustidis}}, \citenamefont {{Bacon}}, \citenamefont {{Baker}},\ and\ \citenamefont {et~al.}}]{Amendola2018:Euclid}%
  \BibitemOpen
  \bibfield  {author} {\bibinfo {author} {\bibfnamefont {L.}~\bibnamefont {{Amendola}}}, \bibinfo {author} {\bibfnamefont {S.}~\bibnamefont {{Appleby}}}, \bibinfo {author} {\bibfnamefont {A.}~\bibnamefont {{Avgoustidis}}}, \bibinfo {author} {\bibfnamefont {D.}~\bibnamefont {{Bacon}}}, \bibinfo {author} {\bibfnamefont {T.}~\bibnamefont {{Baker}}},\ and\ \bibinfo {author} {\bibnamefont {et~al.}},\ }\href {https://doi.org/10.1007/s41114-017-0010-3} {\bibfield  {journal} {\bibinfo  {journal} {Living Reviews in Relativity}\ }\textbf {\bibinfo {volume} {21}},\ \bibinfo {eid} {2} (\bibinfo {year} {2018})},\ \Eprint {https://arxiv.org/abs/1606.00180} {arXiv:1606.00180 [astro-ph.CO]} \BibitemShut {NoStop}%
\bibitem [{\citenamefont {{The LSST Dark Energy Science Collaboration}}(2018)}]{LSST2018:DESC}%
  \BibitemOpen
  \bibfield  {author} {\bibinfo {author} {\bibnamefont {{The LSST Dark Energy Science Collaboration}}},\ }\href {https://doi.org/10.48550/arXiv.1809.01669} {\bibfield  {journal} {\bibinfo  {journal} {arXiv e-prints}\ ,\ \bibinfo {eid} {arXiv:1809.01669}} (\bibinfo {year} {2018})},\ \Eprint {https://arxiv.org/abs/1809.01669} {arXiv:1809.01669 [astro-ph.CO]} \BibitemShut {NoStop}%
\bibitem [{\citenamefont {{Eifler}}\ \emph {et~al.}(2021)\citenamefont {{Eifler}}, \citenamefont {{Miyatake}}, \citenamefont {{Krause}}, \citenamefont {{Heinrich}}, \citenamefont {{Miranda}},\ and\ \citenamefont {et~al.}}]{Eifler2021:Roman}%
  \BibitemOpen
  \bibfield  {author} {\bibinfo {author} {\bibfnamefont {T.}~\bibnamefont {{Eifler}}}, \bibinfo {author} {\bibfnamefont {H.}~\bibnamefont {{Miyatake}}}, \bibinfo {author} {\bibfnamefont {E.}~\bibnamefont {{Krause}}}, \bibinfo {author} {\bibfnamefont {C.}~\bibnamefont {{Heinrich}}}, \bibinfo {author} {\bibfnamefont {V.}~\bibnamefont {{Miranda}}},\ and\ \bibinfo {author} {\bibnamefont {et~al.}},\ }\href {https://doi.org/10.1093/mnras/stab1762} {\bibfield  {journal} {\bibinfo  {journal} {\mnras}\ }\textbf {\bibinfo {volume} {507}},\ \bibinfo {pages} {1746} (\bibinfo {year} {2021})},\ \Eprint {https://arxiv.org/abs/2004.05271} {arXiv:2004.05271 [astro-ph.CO]} \BibitemShut {NoStop}%
\bibitem [{\citenamefont {{Troxel}}\ and\ \citenamefont {{Ishak}}(2015)}]{Troxel2015:IA}%
  \BibitemOpen
  \bibfield  {author} {\bibinfo {author} {\bibfnamefont {M.~A.}\ \bibnamefont {{Troxel}}}\ and\ \bibinfo {author} {\bibfnamefont {M.}~\bibnamefont {{Ishak}}},\ }\href {https://doi.org/10.1016/j.physrep.2014.11.001} {\bibfield  {journal} {\bibinfo  {journal} {\physrep}\ }\textbf {\bibinfo {volume} {558}},\ \bibinfo {pages} {1} (\bibinfo {year} {2015})},\ \Eprint {https://arxiv.org/abs/1407.6990} {arXiv:1407.6990 [astro-ph.CO]} \BibitemShut {NoStop}%
\bibitem [{\citenamefont {{Krause}}\ \emph {et~al.}(2016)\citenamefont {{Krause}}, \citenamefont {{Eifler}},\ and\ \citenamefont {{Blazek}}}]{Krause2016:IA}%
  \BibitemOpen
  \bibfield  {author} {\bibinfo {author} {\bibfnamefont {E.}~\bibnamefont {{Krause}}}, \bibinfo {author} {\bibfnamefont {T.}~\bibnamefont {{Eifler}}},\ and\ \bibinfo {author} {\bibfnamefont {J.}~\bibnamefont {{Blazek}}},\ }\href {https://doi.org/10.1093/mnras/stv2615} {\bibfield  {journal} {\bibinfo  {journal} {\mnras}\ }\textbf {\bibinfo {volume} {456}},\ \bibinfo {pages} {207} (\bibinfo {year} {2016})},\ \Eprint {https://arxiv.org/abs/1506.08730} {arXiv:1506.08730 [astro-ph.CO]} \BibitemShut {NoStop}%
\bibitem [{\citenamefont {{Blain}}(2002)}]{Blain2002:2DspecWL}%
  \BibitemOpen
  \bibfield  {author} {\bibinfo {author} {\bibfnamefont {A.~W.}\ \bibnamefont {{Blain}}},\ }\href {https://doi.org/10.1086/341103} {\bibfield  {journal} {\bibinfo  {journal} {\apjl}\ }\textbf {\bibinfo {volume} {570}},\ \bibinfo {pages} {L51} (\bibinfo {year} {2002})},\ \Eprint {https://arxiv.org/abs/astro-ph/0204138} {arXiv:astro-ph/0204138 [astro-ph]} \BibitemShut {NoStop}%
\bibitem [{\citenamefont {{Wittman}}\ and\ \citenamefont {{Self}}(2021)}]{Wittman2021:KL}%
  \BibitemOpen
  \bibfield  {author} {\bibinfo {author} {\bibfnamefont {D.}~\bibnamefont {{Wittman}}}\ and\ \bibinfo {author} {\bibfnamefont {M.}~\bibnamefont {{Self}}},\ }\href {https://doi.org/10.3847/1538-4357/abd548} {\bibfield  {journal} {\bibinfo  {journal} {\apj}\ }\textbf {\bibinfo {volume} {908}},\ \bibinfo {eid} {34} (\bibinfo {year} {2021})}\BibitemShut {NoStop}%
\bibitem [{\citenamefont {{Gurri}}\ \emph {et~al.}(2020)\citenamefont {{Gurri}}, \citenamefont {{Taylor}},\ and\ \citenamefont {{Fluke}}}]{Gurri2020:firstKL}%
  \BibitemOpen
  \bibfield  {author} {\bibinfo {author} {\bibfnamefont {P.}~\bibnamefont {{Gurri}}}, \bibinfo {author} {\bibfnamefont {E.~N.}\ \bibnamefont {{Taylor}}},\ and\ \bibinfo {author} {\bibfnamefont {C.~J.}\ \bibnamefont {{Fluke}}},\ }\href {https://doi.org/10.1093/mnras/staa2893} {\bibfield  {journal} {\bibinfo  {journal} {\mnras}\ }\textbf {\bibinfo {volume} {499}},\ \bibinfo {pages} {4591} (\bibinfo {year} {2020})},\ \Eprint {https://arxiv.org/abs/2009.10067} {arXiv:2009.10067 [astro-ph.GA]} \BibitemShut {NoStop}%
\bibitem [{\citenamefont {{Huff}}\ \emph {et~al.}(2013)\citenamefont {{Huff}}, \citenamefont {{Krause}}, \citenamefont {{Eifler}}, \citenamefont {{Fang}}, \citenamefont {{George}},\ and\ \citenamefont {{Schlegel}}}]{Huff2013:KL}%
  \BibitemOpen
  \bibfield  {author} {\bibinfo {author} {\bibfnamefont {E.~M.}\ \bibnamefont {{Huff}}}, \bibinfo {author} {\bibfnamefont {E.}~\bibnamefont {{Krause}}}, \bibinfo {author} {\bibfnamefont {T.}~\bibnamefont {{Eifler}}}, \bibinfo {author} {\bibfnamefont {X.}~\bibnamefont {{Fang}}}, \bibinfo {author} {\bibfnamefont {M.~R.}\ \bibnamefont {{George}}},\ and\ \bibinfo {author} {\bibfnamefont {D.}~\bibnamefont {{Schlegel}}},\ }\href@noop {} {\bibfield  {journal} {\bibinfo  {journal} {arXiv e-prints}\ ,\ \bibinfo {eid} {arXiv:1311.1489}} (\bibinfo {year} {2013})},\ \Eprint {https://arxiv.org/abs/1311.1489} {arXiv:1311.1489 [astro-ph.CO]} \BibitemShut {NoStop}%
\bibitem [{\citenamefont {{Tully}}\ and\ \citenamefont {{Fisher}}(1977)}]{TF1977}%
  \BibitemOpen
  \bibfield  {author} {\bibinfo {author} {\bibfnamefont {R.~B.}\ \bibnamefont {{Tully}}}\ and\ \bibinfo {author} {\bibfnamefont {J.~R.}\ \bibnamefont {{Fisher}}},\ }\href@noop {} {\bibfield  {journal} {\bibinfo  {journal} {\aap}\ }\textbf {\bibinfo {volume} {54}},\ \bibinfo {pages} {661} (\bibinfo {year} {1977})}\BibitemShut {NoStop}%
\bibitem [{\citenamefont {{R.~S.}}\ \emph {et~al.}(2023)\citenamefont {{R.~S.}}, \citenamefont {{Krause}}, \citenamefont {{Huang}}, \citenamefont {{Huff}}, \citenamefont {{Xu}}, \citenamefont {{Eifler}},\ and\ \citenamefont {{Everett}}}]{RS2023:realistic}%
  \BibitemOpen
  \bibfield  {author} {\bibinfo {author} {\bibfnamefont {P.}~\bibnamefont {{R.~S.}}}, \bibinfo {author} {\bibfnamefont {E.}~\bibnamefont {{Krause}}}, \bibinfo {author} {\bibfnamefont {H.-J.}\ \bibnamefont {{Huang}}}, \bibinfo {author} {\bibfnamefont {E.}~\bibnamefont {{Huff}}}, \bibinfo {author} {\bibfnamefont {J.}~\bibnamefont {{Xu}}}, \bibinfo {author} {\bibfnamefont {T.}~\bibnamefont {{Eifler}}},\ and\ \bibinfo {author} {\bibfnamefont {S.}~\bibnamefont {{Everett}}},\ }\href {https://doi.org/10.1093/mnras/stad2014} {\bibfield  {journal} {\bibinfo  {journal} {\mnras}\ }\textbf {\bibinfo {volume} {524}},\ \bibinfo {pages} {3324} (\bibinfo {year} {2023})},\ \Eprint {https://arxiv.org/abs/2209.11811} {arXiv:2209.11811 [astro-ph.GA]} \BibitemShut {NoStop}%
\bibitem [{\citenamefont {{Xu}}\ \emph {et~al.}(2023)\citenamefont {{Xu}}, \citenamefont {{Eifler}}, \citenamefont {{Huff}}, \citenamefont {{R.~S.}}, \citenamefont {{Huang}}, \citenamefont {{Everett}},\ and\ \citenamefont {{Krause}}}]{Xu2023:Roman}%
  \BibitemOpen
  \bibfield  {author} {\bibinfo {author} {\bibfnamefont {J.}~\bibnamefont {{Xu}}}, \bibinfo {author} {\bibfnamefont {T.}~\bibnamefont {{Eifler}}}, \bibinfo {author} {\bibfnamefont {E.}~\bibnamefont {{Huff}}}, \bibinfo {author} {\bibfnamefont {P.}~\bibnamefont {{R.~S.}}}, \bibinfo {author} {\bibfnamefont {H.-J.}\ \bibnamefont {{Huang}}}, \bibinfo {author} {\bibfnamefont {S.}~\bibnamefont {{Everett}}},\ and\ \bibinfo {author} {\bibfnamefont {E.}~\bibnamefont {{Krause}}},\ }\href {https://doi.org/10.1093/mnras/stac3685} {\bibfield  {journal} {\bibinfo  {journal} {\mnras}\ }\textbf {\bibinfo {volume} {519}},\ \bibinfo {pages} {2535} (\bibinfo {year} {2023})},\ \Eprint {https://arxiv.org/abs/2201.00739} {arXiv:2201.00739 [astro-ph.CO]} \BibitemShut {NoStop}%
\bibitem [{\citenamefont {{Johnston}}\ \emph {et~al.}(2023)\citenamefont {{Johnston}}, \citenamefont {{Westbeek}}, \citenamefont {{Weide}}, \citenamefont {{Chisari}}, \citenamefont {{Dubois}},\ and\ \citenamefont {et~al.}}]{Johnston2023:size}%
  \BibitemOpen
  \bibfield  {author} {\bibinfo {author} {\bibfnamefont {H.}~\bibnamefont {{Johnston}}}, \bibinfo {author} {\bibfnamefont {D.~S.}\ \bibnamefont {{Westbeek}}}, \bibinfo {author} {\bibfnamefont {S.}~\bibnamefont {{Weide}}}, \bibinfo {author} {\bibfnamefont {N.~E.}\ \bibnamefont {{Chisari}}}, \bibinfo {author} {\bibfnamefont {Y.}~\bibnamefont {{Dubois}}},\ and\ \bibinfo {author} {\bibnamefont {et~al.}},\ }\href {https://doi.org/10.1093/mnras/stad201} {\bibfield  {journal} {\bibinfo  {journal} {\mnras}\ }\textbf {\bibinfo {volume} {520}},\ \bibinfo {pages} {1541} (\bibinfo {year} {2023})},\ \Eprint {https://arxiv.org/abs/2209.11063} {arXiv:2209.11063 [astro-ph.CO]} \BibitemShut {NoStop}%
\bibitem [{\citenamefont {{Pelliccia}}\ \emph {et~al.}(2019)\citenamefont {{Pelliccia}}, \citenamefont {{Lemaux}}, \citenamefont {{Tomczak}}, \citenamefont {{Lubin}}, \citenamefont {{Shen}},\ and\ \citenamefont {et~al.}}]{Pelliccia:2019}%
  \BibitemOpen
  \bibfield  {author} {\bibinfo {author} {\bibfnamefont {D.}~\bibnamefont {{Pelliccia}}}, \bibinfo {author} {\bibfnamefont {B.~C.}\ \bibnamefont {{Lemaux}}}, \bibinfo {author} {\bibfnamefont {A.~R.}\ \bibnamefont {{Tomczak}}}, \bibinfo {author} {\bibfnamefont {L.~M.}\ \bibnamefont {{Lubin}}}, \bibinfo {author} {\bibfnamefont {L.}~\bibnamefont {{Shen}}},\ and\ \bibinfo {author} {\bibnamefont {et~al.}},\ }\href {https://doi.org/10.1093/mnras/sty2876} {\bibfield  {journal} {\bibinfo  {journal} {\mnras}\ }\textbf {\bibinfo {volume} {482}},\ \bibinfo {pages} {3514} (\bibinfo {year} {2019})},\ \Eprint {https://arxiv.org/abs/1807.04763} {arXiv:1807.04763 [astro-ph.GA]} \BibitemShut {NoStop}%
\bibitem [{\citenamefont {{P{\'e}rez-Mart{\'\i}nez}}\ \emph {et~al.}(2021)\citenamefont {{P{\'e}rez-Mart{\'\i}nez}}, \citenamefont {{Ziegler}}, \citenamefont {{Dannerbauer}}, \citenamefont {{B{\"o}hm}}, \citenamefont {{Verdugo}}, \citenamefont {{D{\'\i}az}},\ and\ \citenamefont {{Hoyos}}}]{Perez-Mart:2021}%
  \BibitemOpen
  \bibfield  {author} {\bibinfo {author} {\bibfnamefont {J.~M.}\ \bibnamefont {{P{\'e}rez-Mart{\'\i}nez}}}, \bibinfo {author} {\bibfnamefont {B.}~\bibnamefont {{Ziegler}}}, \bibinfo {author} {\bibfnamefont {H.}~\bibnamefont {{Dannerbauer}}}, \bibinfo {author} {\bibfnamefont {A.}~\bibnamefont {{B{\"o}hm}}}, \bibinfo {author} {\bibfnamefont {M.}~\bibnamefont {{Verdugo}}}, \bibinfo {author} {\bibfnamefont {A.~I.}\ \bibnamefont {{D{\'\i}az}}},\ and\ \bibinfo {author} {\bibfnamefont {C.}~\bibnamefont {{Hoyos}}},\ }\href {https://doi.org/10.1051/0004-6361/201936456} {\bibfield  {journal} {\bibinfo  {journal} {\aap}\ }\textbf {\bibinfo {volume} {646}},\ \bibinfo {eid} {A53} (\bibinfo {year} {2021})},\ \Eprint {https://arxiv.org/abs/2007.13068} {arXiv:2007.13068 [astro-ph.GA]} \BibitemShut {NoStop}%
\bibitem [{\citenamefont {{Abril-Melgarejo}}\ \emph {et~al.}(2021)\citenamefont {{Abril-Melgarejo}}, \citenamefont {{Epinat}}, \citenamefont {{Mercier}}, \citenamefont {{Contini}},\ and\ \citenamefont {et~al.}}]{Abril-Melgarejo:2021}%
  \BibitemOpen
  \bibfield  {author} {\bibinfo {author} {\bibfnamefont {V.}~\bibnamefont {{Abril-Melgarejo}}}, \bibinfo {author} {\bibfnamefont {B.}~\bibnamefont {{Epinat}}}, \bibinfo {author} {\bibfnamefont {W.}~\bibnamefont {{Mercier}}}, \bibinfo {author} {\bibfnamefont {T.}~\bibnamefont {{Contini}}},\ and\ \bibinfo {author} {\bibnamefont {et~al.}},\ }\href {https://doi.org/10.1051/0004-6361/202038818} {\bibfield  {journal} {\bibinfo  {journal} {\aap}\ }\textbf {\bibinfo {volume} {647}},\ \bibinfo {eid} {A152} (\bibinfo {year} {2021})},\ \Eprint {https://arxiv.org/abs/2101.08069} {arXiv:2101.08069 [astro-ph.GA]} \BibitemShut {NoStop}%
\bibitem [{\citenamefont {{Mercier}}\ \emph {et~al.}(2022)\citenamefont {{Mercier}}, \citenamefont {{Epinat}}, \citenamefont {{Contini}}, \citenamefont {{Abril-Melgarejo}}, \citenamefont {{Boogaard}},\ and\ \citenamefont {et~al.}}]{Mercier:2022}%
  \BibitemOpen
  \bibfield  {author} {\bibinfo {author} {\bibfnamefont {W.}~\bibnamefont {{Mercier}}}, \bibinfo {author} {\bibfnamefont {B.}~\bibnamefont {{Epinat}}}, \bibinfo {author} {\bibfnamefont {T.}~\bibnamefont {{Contini}}}, \bibinfo {author} {\bibfnamefont {V.}~\bibnamefont {{Abril-Melgarejo}}}, \bibinfo {author} {\bibfnamefont {L.}~\bibnamefont {{Boogaard}}},\ and\ \bibinfo {author} {\bibnamefont {et~al.}},\ }\href {https://doi.org/10.1051/0004-6361/202243110} {\bibfield  {journal} {\bibinfo  {journal} {\aap}\ }\textbf {\bibinfo {volume} {665}},\ \bibinfo {eid} {A54} (\bibinfo {year} {2022})},\ \Eprint {https://arxiv.org/abs/2204.08724} {arXiv:2204.08724 [astro-ph.GA]} \BibitemShut {NoStop}%
\bibitem [{\citenamefont {{Blazek}}\ \emph {et~al.}(2019)\citenamefont {{Blazek}}, \citenamefont {{MacCrann}}, \citenamefont {{Troxel}},\ and\ \citenamefont {{Fang}}}]{Blazek2019:IA}%
  \BibitemOpen
  \bibfield  {author} {\bibinfo {author} {\bibfnamefont {J.~A.}\ \bibnamefont {{Blazek}}}, \bibinfo {author} {\bibfnamefont {N.}~\bibnamefont {{MacCrann}}}, \bibinfo {author} {\bibfnamefont {M.~A.}\ \bibnamefont {{Troxel}}},\ and\ \bibinfo {author} {\bibfnamefont {X.}~\bibnamefont {{Fang}}},\ }\href {https://doi.org/10.1103/PhysRevD.100.103506} {\bibfield  {journal} {\bibinfo  {journal} {\prd}\ }\textbf {\bibinfo {volume} {100}},\ \bibinfo {eid} {103506} (\bibinfo {year} {2019})},\ \Eprint {https://arxiv.org/abs/1708.09247} {arXiv:1708.09247 [astro-ph.CO]} \BibitemShut {NoStop}%
\bibitem [{\citenamefont {{Tonegawa}}\ \emph {et~al.}(2018)\citenamefont {{Tonegawa}}, \citenamefont {{Okumura}}, \citenamefont {{Totani}}, \citenamefont {{Dalton}}, \citenamefont {{Glazebrook}},\ and\ \citenamefont {{Yabe}}}]{Tonegawa2018:IA}%
  \BibitemOpen
  \bibfield  {author} {\bibinfo {author} {\bibfnamefont {M.}~\bibnamefont {{Tonegawa}}}, \bibinfo {author} {\bibfnamefont {T.}~\bibnamefont {{Okumura}}}, \bibinfo {author} {\bibfnamefont {T.}~\bibnamefont {{Totani}}}, \bibinfo {author} {\bibfnamefont {G.}~\bibnamefont {{Dalton}}}, \bibinfo {author} {\bibfnamefont {K.}~\bibnamefont {{Glazebrook}}},\ and\ \bibinfo {author} {\bibfnamefont {K.}~\bibnamefont {{Yabe}}},\ }\href {https://doi.org/10.1093/pasj/psy030} {\bibfield  {journal} {\bibinfo  {journal} {\pasj}\ }\textbf {\bibinfo {volume} {70}},\ \bibinfo {eid} {41} (\bibinfo {year} {2018})},\ \Eprint {https://arxiv.org/abs/1708.02224} {arXiv:1708.02224 [astro-ph.CO]} \BibitemShut {NoStop}%
\bibitem [{\citenamefont {{Samuroff}}\ \emph {et~al.}(2023)\citenamefont {{Samuroff}}, \citenamefont {{Mandelbaum}}, \citenamefont {{Blazek}}, \citenamefont {{Campos}}, \citenamefont {{MacCrann}},\ and\ \citenamefont {{DES Collaboration}}}]{Samuroff2023:IA}%
  \BibitemOpen
  \bibfield  {author} {\bibinfo {author} {\bibfnamefont {S.}~\bibnamefont {{Samuroff}}}, \bibinfo {author} {\bibfnamefont {R.}~\bibnamefont {{Mandelbaum}}}, \bibinfo {author} {\bibfnamefont {J.}~\bibnamefont {{Blazek}}}, \bibinfo {author} {\bibfnamefont {A.}~\bibnamefont {{Campos}}}, \bibinfo {author} {\bibfnamefont {N.}~\bibnamefont {{MacCrann}}},\ and\ \bibinfo {author} {\bibnamefont {{DES Collaboration}}},\ }\href {https://doi.org/10.1093/mnras/stad2013} {\bibfield  {journal} {\bibinfo  {journal} {\mnras}\ }\textbf {\bibinfo {volume} {524}},\ \bibinfo {pages} {2195} (\bibinfo {year} {2023})},\ \Eprint {https://arxiv.org/abs/2212.11319} {arXiv:2212.11319 [astro-ph.CO]} \BibitemShut {NoStop}%
\bibitem [{\citenamefont {{To}}\ \emph {et~al.}(2024)\citenamefont {{To}}, \citenamefont {{Pandey}}, \citenamefont {{Krause}}, \citenamefont {{Dalal}}, \citenamefont {{Anbajagane}},\ and\ \citenamefont {{Weinberg}}}]{To2024:smallscale}%
  \BibitemOpen
  \bibfield  {author} {\bibinfo {author} {\bibfnamefont {C.-H.}\ \bibnamefont {{To}}}, \bibinfo {author} {\bibfnamefont {S.}~\bibnamefont {{Pandey}}}, \bibinfo {author} {\bibfnamefont {E.}~\bibnamefont {{Krause}}}, \bibinfo {author} {\bibfnamefont {N.}~\bibnamefont {{Dalal}}}, \bibinfo {author} {\bibfnamefont {D.}~\bibnamefont {{Anbajagane}}},\ and\ \bibinfo {author} {\bibfnamefont {D.~H.}\ \bibnamefont {{Weinberg}}},\ }\href {https://doi.org/10.48550/arXiv.2402.00110} {\bibfield  {journal} {\bibinfo  {journal} {arXiv e-prints}\ ,\ \bibinfo {eid} {arXiv:2402.00110}} (\bibinfo {year} {2024})},\ \Eprint {https://arxiv.org/abs/2402.00110} {arXiv:2402.00110 [astro-ph.CO]} \BibitemShut {NoStop}%
\bibitem [{\citenamefont {{Nelson}}\ \emph {et~al.}(2019)\citenamefont {{Nelson}}, \citenamefont {{Springel}}, \citenamefont {{Pillepich}}, \citenamefont {{Rodriguez-Gomez}}, \citenamefont {{Torrey}},\ and\ \citenamefont {et~al.}}]{Nelson2019:TNGsum}%
  \BibitemOpen
  \bibfield  {author} {\bibinfo {author} {\bibfnamefont {D.}~\bibnamefont {{Nelson}}}, \bibinfo {author} {\bibfnamefont {V.}~\bibnamefont {{Springel}}}, \bibinfo {author} {\bibfnamefont {A.}~\bibnamefont {{Pillepich}}}, \bibinfo {author} {\bibfnamefont {V.}~\bibnamefont {{Rodriguez-Gomez}}}, \bibinfo {author} {\bibfnamefont {P.}~\bibnamefont {{Torrey}}},\ and\ \bibinfo {author} {\bibnamefont {et~al.}},\ }\href {https://doi.org/10.1186/s40668-019-0028-x} {\bibfield  {journal} {\bibinfo  {journal} {Computational Astrophysics and Cosmology}\ }\textbf {\bibinfo {volume} {6}},\ \bibinfo {eid} {2} (\bibinfo {year} {2019})},\ \Eprint {https://arxiv.org/abs/1812.05609} {arXiv:1812.05609 [astro-ph.GA]} \BibitemShut {NoStop}%
\bibitem [{\citenamefont {{Pillepich}}\ \emph {et~al.}(2018)\citenamefont {{Pillepich}}, \citenamefont {{Nelson}}, \citenamefont {{Hernquist}}, \citenamefont {{Springel}}, \citenamefont {{Pakmor}},\ and\ \citenamefont {et~al.}}]{Piilepich2018:TNGstar}%
  \BibitemOpen
  \bibfield  {author} {\bibinfo {author} {\bibfnamefont {A.}~\bibnamefont {{Pillepich}}}, \bibinfo {author} {\bibfnamefont {D.}~\bibnamefont {{Nelson}}}, \bibinfo {author} {\bibfnamefont {L.}~\bibnamefont {{Hernquist}}}, \bibinfo {author} {\bibfnamefont {V.}~\bibnamefont {{Springel}}}, \bibinfo {author} {\bibfnamefont {R.}~\bibnamefont {{Pakmor}}},\ and\ \bibinfo {author} {\bibnamefont {et~al.}},\ }\href {https://doi.org/10.1093/mnras/stx3112} {\bibfield  {journal} {\bibinfo  {journal} {\mnras}\ }\textbf {\bibinfo {volume} {475}},\ \bibinfo {pages} {648} (\bibinfo {year} {2018})},\ \Eprint {https://arxiv.org/abs/1707.03406} {arXiv:1707.03406 [astro-ph.GA]} \BibitemShut {NoStop}%
\bibitem [{\citenamefont {{Nelson}}\ \emph {et~al.}(2018)\citenamefont {{Nelson}}, \citenamefont {{Pillepich}}, \citenamefont {{Springel}}, \citenamefont {{Weinberger}}, \citenamefont {{Hernquist}},\ and\ \citenamefont {et~al.}}]{Nelson2018:TNGcolor}%
  \BibitemOpen
  \bibfield  {author} {\bibinfo {author} {\bibfnamefont {D.}~\bibnamefont {{Nelson}}}, \bibinfo {author} {\bibfnamefont {A.}~\bibnamefont {{Pillepich}}}, \bibinfo {author} {\bibfnamefont {V.}~\bibnamefont {{Springel}}}, \bibinfo {author} {\bibfnamefont {R.}~\bibnamefont {{Weinberger}}}, \bibinfo {author} {\bibfnamefont {L.}~\bibnamefont {{Hernquist}}},\ and\ \bibinfo {author} {\bibnamefont {et~al.}},\ }\href {https://doi.org/10.1093/mnras/stx3040} {\bibfield  {journal} {\bibinfo  {journal} {\mnras}\ }\textbf {\bibinfo {volume} {475}},\ \bibinfo {pages} {624} (\bibinfo {year} {2018})},\ \Eprint {https://arxiv.org/abs/1707.03395} {arXiv:1707.03395 [astro-ph.GA]} \BibitemShut {NoStop}%
\bibitem [{\citenamefont {{Naiman}}\ \emph {et~al.}(2018)\citenamefont {{Naiman}}, \citenamefont {{Pillepich}}, \citenamefont {{Springel}}, \citenamefont {{Ramirez-Ruiz}}, \citenamefont {{Torrey}},\ and\ \citenamefont {et~al.}}]{Naiman2018:TNGchem}%
  \BibitemOpen
  \bibfield  {author} {\bibinfo {author} {\bibfnamefont {J.~P.}\ \bibnamefont {{Naiman}}}, \bibinfo {author} {\bibfnamefont {A.}~\bibnamefont {{Pillepich}}}, \bibinfo {author} {\bibfnamefont {V.}~\bibnamefont {{Springel}}}, \bibinfo {author} {\bibfnamefont {E.}~\bibnamefont {{Ramirez-Ruiz}}}, \bibinfo {author} {\bibfnamefont {P.}~\bibnamefont {{Torrey}}},\ and\ \bibinfo {author} {\bibnamefont {et~al.}},\ }\href {https://doi.org/10.1093/mnras/sty618} {\bibfield  {journal} {\bibinfo  {journal} {\mnras}\ }\textbf {\bibinfo {volume} {477}},\ \bibinfo {pages} {1206} (\bibinfo {year} {2018})},\ \Eprint {https://arxiv.org/abs/1707.03401} {arXiv:1707.03401 [astro-ph.GA]} \BibitemShut {NoStop}%
\bibitem [{\citenamefont {{Springel}}\ \emph {et~al.}(2018)\citenamefont {{Springel}}, \citenamefont {{Pakmor}}, \citenamefont {{Pillepich}}, \citenamefont {{Weinberger}}, \citenamefont {{Nelson}},\ and\ \citenamefont {et~al.}}]{Springel2018:TNGcluster}%
  \BibitemOpen
  \bibfield  {author} {\bibinfo {author} {\bibfnamefont {V.}~\bibnamefont {{Springel}}}, \bibinfo {author} {\bibfnamefont {R.}~\bibnamefont {{Pakmor}}}, \bibinfo {author} {\bibfnamefont {A.}~\bibnamefont {{Pillepich}}}, \bibinfo {author} {\bibfnamefont {R.}~\bibnamefont {{Weinberger}}}, \bibinfo {author} {\bibfnamefont {D.}~\bibnamefont {{Nelson}}},\ and\ \bibinfo {author} {\bibnamefont {et~al.}},\ }\href {https://doi.org/10.1093/mnras/stx3304} {\bibfield  {journal} {\bibinfo  {journal} {\mnras}\ }\textbf {\bibinfo {volume} {475}},\ \bibinfo {pages} {676} (\bibinfo {year} {2018})},\ \Eprint {https://arxiv.org/abs/1707.03397} {arXiv:1707.03397 [astro-ph.GA]} \BibitemShut {NoStop}%
\bibitem [{\citenamefont {{Marinacci}}\ \emph {et~al.}(2018)\citenamefont {{Marinacci}}, \citenamefont {{Vogelsberger}}, \citenamefont {{Pakmor}}, \citenamefont {{Torrey}}, \citenamefont {{Springel}},\ and\ \citenamefont {et~al.}}]{Marinacci2018:TNGmag}%
  \BibitemOpen
  \bibfield  {author} {\bibinfo {author} {\bibfnamefont {F.}~\bibnamefont {{Marinacci}}}, \bibinfo {author} {\bibfnamefont {M.}~\bibnamefont {{Vogelsberger}}}, \bibinfo {author} {\bibfnamefont {R.}~\bibnamefont {{Pakmor}}}, \bibinfo {author} {\bibfnamefont {P.}~\bibnamefont {{Torrey}}}, \bibinfo {author} {\bibfnamefont {V.}~\bibnamefont {{Springel}}},\ and\ \bibinfo {author} {\bibnamefont {et~al.}},\ }\href {https://doi.org/10.1093/mnras/sty2206} {\bibfield  {journal} {\bibinfo  {journal} {\mnras}\ }\textbf {\bibinfo {volume} {480}},\ \bibinfo {pages} {5113} (\bibinfo {year} {2018})},\ \Eprint {https://arxiv.org/abs/1707.03396} {arXiv:1707.03396 [astro-ph.CO]} \BibitemShut {NoStop}%
\bibitem [{\citenamefont {{Planck Collaboration}}(2016)}]{Planck2016}%
  \BibitemOpen
  \bibfield  {author} {\bibinfo {author} {\bibnamefont {{Planck Collaboration}}},\ }\href {https://doi.org/10.1051/0004-6361/201525830} {\bibfield  {journal} {\bibinfo  {journal} {\aap}\ }\textbf {\bibinfo {volume} {594}},\ \bibinfo {eid} {A13} (\bibinfo {year} {2016})},\ \Eprint {https://arxiv.org/abs/1502.01589} {arXiv:1502.01589 [astro-ph.CO]} \BibitemShut {NoStop}%
\bibitem [{\citenamefont {{Sales}}\ \emph {et~al.}(2012)\citenamefont {{Sales}}, \citenamefont {{Navarro}}, \citenamefont {{Theuns}}, \citenamefont {{Schaye}}, \citenamefont {{White}},\ and\ \citenamefont {et~al.}}]{Sales2012:Kappa}%
  \BibitemOpen
  \bibfield  {author} {\bibinfo {author} {\bibfnamefont {L.~V.}\ \bibnamefont {{Sales}}}, \bibinfo {author} {\bibfnamefont {J.~F.}\ \bibnamefont {{Navarro}}}, \bibinfo {author} {\bibfnamefont {T.}~\bibnamefont {{Theuns}}}, \bibinfo {author} {\bibfnamefont {J.}~\bibnamefont {{Schaye}}}, \bibinfo {author} {\bibfnamefont {S.~D.~M.}\ \bibnamefont {{White}}},\ and\ \bibinfo {author} {\bibnamefont {et~al.}},\ }\href {https://doi.org/10.1111/j.1365-2966.2012.20975.x} {\bibfield  {journal} {\bibinfo  {journal} {\mnras}\ }\textbf {\bibinfo {volume} {423}},\ \bibinfo {pages} {1544} (\bibinfo {year} {2012})},\ \Eprint {https://arxiv.org/abs/1112.2220} {arXiv:1112.2220 [astro-ph.CO]} \BibitemShut {NoStop}%
\bibitem [{\citenamefont {{Rodriguez-Gomez}}\ \emph {et~al.}(2017)\citenamefont {{Rodriguez-Gomez}}, \citenamefont {{Sales}}, \citenamefont {{Genel}}, \citenamefont {{Pillepich}}, \citenamefont {{Zjupa}},\ and\ \citenamefont {et~al.}}]{Rodriguez2017:Kappa}%
  \BibitemOpen
  \bibfield  {author} {\bibinfo {author} {\bibfnamefont {V.}~\bibnamefont {{Rodriguez-Gomez}}}, \bibinfo {author} {\bibfnamefont {L.~V.}\ \bibnamefont {{Sales}}}, \bibinfo {author} {\bibfnamefont {S.}~\bibnamefont {{Genel}}}, \bibinfo {author} {\bibfnamefont {A.}~\bibnamefont {{Pillepich}}}, \bibinfo {author} {\bibfnamefont {J.}~\bibnamefont {{Zjupa}}},\ and\ \bibinfo {author} {\bibnamefont {et~al.}},\ }\href {https://doi.org/10.1093/mnras/stx305} {\bibfield  {journal} {\bibinfo  {journal} {\mnras}\ }\textbf {\bibinfo {volume} {467}},\ \bibinfo {pages} {3083} (\bibinfo {year} {2017})},\ \Eprint {https://arxiv.org/abs/1609.09498} {arXiv:1609.09498 [astro-ph.GA]} \BibitemShut {NoStop}%
\bibitem [{\citenamefont {{Zjupa}}\ \emph {et~al.}(2022)\citenamefont {{Zjupa}}, \citenamefont {{Sch{\"a}fer}},\ and\ \citenamefont {{Hahn}}}]{Zjupa2022:selection}%
  \BibitemOpen
  \bibfield  {author} {\bibinfo {author} {\bibfnamefont {J.}~\bibnamefont {{Zjupa}}}, \bibinfo {author} {\bibfnamefont {B.~M.}\ \bibnamefont {{Sch{\"a}fer}}},\ and\ \bibinfo {author} {\bibfnamefont {O.}~\bibnamefont {{Hahn}}},\ }\href {https://doi.org/10.1093/mnras/stac042} {\bibfield  {journal} {\bibinfo  {journal} {\mnras}\ }\textbf {\bibinfo {volume} {514}},\ \bibinfo {pages} {2049} (\bibinfo {year} {2022})}\BibitemShut {NoStop}%
\bibitem [{\citenamefont {{Pillepich}}\ \emph {et~al.}(2019)\citenamefont {{Pillepich}}, \citenamefont {{Nelson}}, \citenamefont {{Springel}}, \citenamefont {{Pakmor}}, \citenamefont {{Torrey}},\ and\ \citenamefont {et~al.}}]{Pillepich2019:TNGdisk}%
  \BibitemOpen
  \bibfield  {author} {\bibinfo {author} {\bibfnamefont {A.}~\bibnamefont {{Pillepich}}}, \bibinfo {author} {\bibfnamefont {D.}~\bibnamefont {{Nelson}}}, \bibinfo {author} {\bibfnamefont {V.}~\bibnamefont {{Springel}}}, \bibinfo {author} {\bibfnamefont {R.}~\bibnamefont {{Pakmor}}}, \bibinfo {author} {\bibfnamefont {P.}~\bibnamefont {{Torrey}}},\ and\ \bibinfo {author} {\bibnamefont {et~al.}},\ }\href {https://doi.org/10.1093/mnras/stz2338} {\bibfield  {journal} {\bibinfo  {journal} {\mnras}\ }\textbf {\bibinfo {volume} {490}},\ \bibinfo {pages} {3196} (\bibinfo {year} {2019})},\ \Eprint {https://arxiv.org/abs/1902.05553} {arXiv:1902.05553 [astro-ph.GA]} \BibitemShut {NoStop}%
\bibitem [{\citenamefont {{Kennicutt}}(1998)}]{Kennicutt:1998}%
  \BibitemOpen
  \bibfield  {author} {\bibinfo {author} {\bibfnamefont {J.}~\bibnamefont {{Kennicutt}}, \bibfnamefont {Robert~C.}},\ }\href {https://doi.org/10.1146/annurev.astro.36.1.189} {\bibfield  {journal} {\bibinfo  {journal} {\araa}\ }\textbf {\bibinfo {volume} {36}},\ \bibinfo {pages} {189} (\bibinfo {year} {1998})},\ \Eprint {https://arxiv.org/abs/astro-ph/9807187} {arXiv:astro-ph/9807187 [astro-ph]} \BibitemShut {NoStop}%
\bibitem [{\citenamefont {{Crittenden}}\ \emph {et~al.}(2001)\citenamefont {{Crittenden}}, \citenamefont {{Natarajan}}, \citenamefont {{Pen}},\ and\ \citenamefont {{Theuns}}}]{Crittenden:2001ellip}%
  \BibitemOpen
  \bibfield  {author} {\bibinfo {author} {\bibfnamefont {R.~G.}\ \bibnamefont {{Crittenden}}}, \bibinfo {author} {\bibfnamefont {P.}~\bibnamefont {{Natarajan}}}, \bibinfo {author} {\bibfnamefont {U.-L.}\ \bibnamefont {{Pen}}},\ and\ \bibinfo {author} {\bibfnamefont {T.}~\bibnamefont {{Theuns}}},\ }\href {https://doi.org/10.1086/322370} {\bibfield  {journal} {\bibinfo  {journal} {\apj}\ }\textbf {\bibinfo {volume} {559}},\ \bibinfo {pages} {552} (\bibinfo {year} {2001})},\ \Eprint {https://arxiv.org/abs/astro-ph/0009052} {arXiv:astro-ph/0009052 [astro-ph]} \BibitemShut {NoStop}%
\bibitem [{\citenamefont {Heavens}\ and\ \citenamefont {Peacock}(1988)}]{Heavens1988:tensor_shear}%
  \BibitemOpen
  \bibfield  {author} {\bibinfo {author} {\bibfnamefont {A.}~\bibnamefont {Heavens}}\ and\ \bibinfo {author} {\bibfnamefont {J.}~\bibnamefont {Peacock}},\ }\href {https://doi.org/10.1093/mnras/232.2.339} {\bibfield  {journal} {\bibinfo  {journal} {Monthly Notices of the Royal Astronomical Society}\ }\textbf {\bibinfo {volume} {232}},\ \bibinfo {pages} {339} (\bibinfo {year} {1988})},\ \Eprint {https://arxiv.org/abs/https://academic.oup.com/mnras/article-pdf/232/2/339/3316537/mnras232-0339.pdf} {https://academic.oup.com/mnras/article-pdf/232/2/339/3316537/mnras232-0339.pdf} \BibitemShut {NoStop}%
\bibitem [{\citenamefont {{Catelan}}\ and\ \citenamefont {{Theuns}}(1996)}]{Catelan1996:tensor_shear}%
  \BibitemOpen
  \bibfield  {author} {\bibinfo {author} {\bibfnamefont {P.}~\bibnamefont {{Catelan}}}\ and\ \bibinfo {author} {\bibfnamefont {T.}~\bibnamefont {{Theuns}}},\ }\href {https://doi.org/10.1093/mnras/282.2.436} {\bibfield  {journal} {\bibinfo  {journal} {\mnras}\ }\textbf {\bibinfo {volume} {282}},\ \bibinfo {pages} {436} (\bibinfo {year} {1996})},\ \Eprint {https://arxiv.org/abs/astro-ph/9604077} {arXiv:astro-ph/9604077 [astro-ph]} \BibitemShut {NoStop}%
\bibitem [{\citenamefont {{Sheth}}\ and\ \citenamefont {{Tormen}}(2002)}]{Sheth2002:q_diff_from_delta}%
  \BibitemOpen
  \bibfield  {author} {\bibinfo {author} {\bibfnamefont {R.~K.}\ \bibnamefont {{Sheth}}}\ and\ \bibinfo {author} {\bibfnamefont {G.}~\bibnamefont {{Tormen}}},\ }\href {https://doi.org/10.1046/j.1365-8711.2002.04950.x} {\bibfield  {journal} {\bibinfo  {journal} {\mnras}\ }\textbf {\bibinfo {volume} {329}},\ \bibinfo {pages} {61} (\bibinfo {year} {2002})},\ \Eprint {https://arxiv.org/abs/astro-ph/0105113} {arXiv:astro-ph/0105113 [astro-ph]} \BibitemShut {NoStop}%
\bibitem [{\citenamefont {{Jarvis}}\ \emph {et~al.}(2004)\citenamefont {{Jarvis}}, \citenamefont {{Bernstein}},\ and\ \citenamefont {{Jain}}}]{treecorr}%
  \BibitemOpen
  \bibfield  {author} {\bibinfo {author} {\bibfnamefont {M.}~\bibnamefont {{Jarvis}}}, \bibinfo {author} {\bibfnamefont {G.}~\bibnamefont {{Bernstein}}},\ and\ \bibinfo {author} {\bibfnamefont {B.}~\bibnamefont {{Jain}}},\ }\href {https://doi.org/10.1111/j.1365-2966.2004.07926.x} {\bibfield  {journal} {\bibinfo  {journal} {\mnras}\ }\textbf {\bibinfo {volume} {352}},\ \bibinfo {pages} {338} (\bibinfo {year} {2004})},\ \Eprint {https://arxiv.org/abs/astro-ph/0307393} {arXiv:astro-ph/0307393 [astro-ph]} \BibitemShut {NoStop}%
\bibitem [{\citenamefont {{Tinker}}\ \emph {et~al.}(2010)\citenamefont {{Tinker}}, \citenamefont {{Robertson}}, \citenamefont {{Kravtsov}}, \citenamefont {{Klypin}}, \citenamefont {{Warren}},\ and\ \citenamefont {et~al.}}]{Tinker2010:bias}%
  \BibitemOpen
  \bibfield  {author} {\bibinfo {author} {\bibfnamefont {J.~L.}\ \bibnamefont {{Tinker}}}, \bibinfo {author} {\bibfnamefont {B.~E.}\ \bibnamefont {{Robertson}}}, \bibinfo {author} {\bibfnamefont {A.~V.}\ \bibnamefont {{Kravtsov}}}, \bibinfo {author} {\bibfnamefont {A.}~\bibnamefont {{Klypin}}}, \bibinfo {author} {\bibfnamefont {M.~S.}\ \bibnamefont {{Warren}}},\ and\ \bibinfo {author} {\bibnamefont {et~al.}},\ }\href {https://doi.org/10.1088/0004-637X/724/2/878} {\bibfield  {journal} {\bibinfo  {journal} {\apj}\ }\textbf {\bibinfo {volume} {724}},\ \bibinfo {pages} {878} (\bibinfo {year} {2010})},\ \Eprint {https://arxiv.org/abs/1001.3162} {arXiv:1001.3162 [astro-ph.CO]} \BibitemShut {NoStop}%
\bibitem [{\citenamefont {{Dav{\'e}}}\ \emph {et~al.}(2020)\citenamefont {{Dav{\'e}}}, \citenamefont {{Crain}}, \citenamefont {{Stevens}}, \citenamefont {{Narayanan}}, \citenamefont {{Saintonge}},\ and\ \citenamefont {et~al.}}]{Dave2020}%
  \BibitemOpen
  \bibfield  {author} {\bibinfo {author} {\bibfnamefont {R.}~\bibnamefont {{Dav{\'e}}}}, \bibinfo {author} {\bibfnamefont {R.~A.}\ \bibnamefont {{Crain}}}, \bibinfo {author} {\bibfnamefont {A.~R.~H.}\ \bibnamefont {{Stevens}}}, \bibinfo {author} {\bibfnamefont {D.}~\bibnamefont {{Narayanan}}}, \bibinfo {author} {\bibfnamefont {A.}~\bibnamefont {{Saintonge}}},\ and\ \bibinfo {author} {\bibnamefont {et~al.}},\ }\href {https://doi.org/10.1093/mnras/staa1894} {\bibfield  {journal} {\bibinfo  {journal} {\mnras}\ }\textbf {\bibinfo {volume} {497}},\ \bibinfo {pages} {146} (\bibinfo {year} {2020})},\ \Eprint {https://arxiv.org/abs/2002.07226} {arXiv:2002.07226 [astro-ph.GA]} \BibitemShut {NoStop}%
\bibitem [{\citenamefont {{Tenneti}}\ \emph {et~al.}(2016)\citenamefont {{Tenneti}}, \citenamefont {{Mandelbaum}},\ and\ \citenamefont {{Di Matteo}}}]{Tenneti2016:IA}%
  \BibitemOpen
  \bibfield  {author} {\bibinfo {author} {\bibfnamefont {A.}~\bibnamefont {{Tenneti}}}, \bibinfo {author} {\bibfnamefont {R.}~\bibnamefont {{Mandelbaum}}},\ and\ \bibinfo {author} {\bibfnamefont {T.}~\bibnamefont {{Di Matteo}}},\ }\href {https://doi.org/10.1093/mnras/stw1823} {\bibfield  {journal} {\bibinfo  {journal} {\mnras}\ }\textbf {\bibinfo {volume} {462}},\ \bibinfo {pages} {2668} (\bibinfo {year} {2016})},\ \Eprint {https://arxiv.org/abs/1510.07024} {arXiv:1510.07024 [astro-ph.CO]} \BibitemShut {NoStop}%
\end{thebibliography}%

\end{document}